\documentclass[manuscript]{aastex61}

\usepackage{graphicx}
\usepackage{hyperref}
\usepackage{color}
\usepackage{xcolor}

\let\textbf\relax

\newcommand{\teff}{$T_{\rm eff}$}
\newcommand{\logg}{$\log{g}$}
\newcommand{\feh}{[Fe/H]}
\newcommand{\vmic}{$\xi$}
\newcommand{\vsini}{$v\sin i$}
\newcommand{\msun}{$M_\odot$}
\newcommand{\rsun}{$R_\odot$}
\newcommand{\mstar}{$M_\star$}

\newcommand{\obnat}{Observat\'orio Nacional, Rua General Jos\'e Cristino, 77, 20921-400, S\~ao Crist\'ov\~ao, Rio de Janeiro, RJ, Brazil; lghezzi@on.br}
\newcommand{\cfa}{Harvard-Smithsonian Center for Astrophysics, 60 Garden Street, Cambridge, MA 02138 USA}
\newcommand{\chicago}{Department of Astronomy and Astrophysics, University of Chicago, 5640 S. Ellis Ave, Chicago, IL 60637, USA}
\newcommand{\sagan}{Sagan Fellow}

\begin{document}

\author{Luan Ghezzi}
\affiliation{\obnat}

\author{Benjamin~T.~Montet}
\altaffiliation{\sagan}
\affiliation{\chicago}

\author{John Asher Johnson}
\affiliation{\cfa}

\shorttitle{Retired A Stars Revisited}
\shortauthors{Ghezzi, Montet \& Johnson}

\title{Retired A Stars Revisited: An Updated Giant Planet Occurrence Rate as a Function of Stellar Metallicity and Mass}

\begin{abstract}
Exoplanet surveys of evolved stars have  provided increasing evidence that the formation of giant planets depends not only on stellar metallicity ([Fe/H]), but also the mass (\mstar). However, measuring accurate masses for subgiants and giants is far more challenging than it is for their main-sequence counterparts, which has led to recent concerns regarding the veracity of the correlation between stellar mass and planet occurrence. In order to address these concerns we use HIRES spectra to perform a spectroscopic analysis on an  sample of 245 subgiants and derive new atmospheric and physical parameters. We also calculate the space velocities of this sample in a homogeneous manner for the first time. \textbf{When reddening corrections are considered in the calculations of stellar masses and a -0.12 \msun\ offset is applied to the results, the masses of the subgiants are consistent with their space velocity distributions, contrary to claims in the literature. Similarly, our measurement of their rotational velocities provide additional confirmation that the masses of subgiants with $M_\star \geq 1.6$~M$_\sun$ (the ``Retired A Stars'') have not been overestimated in previous analyses.} Using these new results for our sample of evolved stars, together with an updated sample of FGKM dwarfs, we confirm that giant planet occurrence increases with both stellar mass and metallicity up to 2.0~M$_{\odot}$. We show that the probability of formation of a giant planet is approximately a one-to-one function of the total amount of metals in the protoplanetary disk \mstar\,$10^{[Fe/H]}$. This correlation provides additional support for the core accretion mechanism of planet formation.
\end{abstract}



\section{Introduction}
\label{introduction}

The ever increasing sample of exoplanets has facilitated many robust studies of how planet formation and evolution are affected by the physical properties of their host stars. One such result is the correlation between stellar metallicity and the occurrence rate of giant planets around FGK dwarf and subgiant stars (e.g., \citealt{gonzalez97,fv05,ghezzi10a,wang15}). Similar analyses for the stellar masses, however, are much more scarce and controversial. \textbf{The main limitation appears on the massive end ($\gtrsim$1.2 \msun) of the planet search samples, since the detection or confirmation of exoplanets around main-sequence A and early F stars using radial velocities (RVs) is more difficult due to their higher effective temperatures and larger rotational velocities.}

In order to overcome this issue, radial velocity surveys have focused on intermediate-mass stars. As stars evolve off of the main sequence towards the red giant branch (RGB), their effective temperatures decrease and their rotation velocities slow down, resulting in spectra from which precise Doppler shifts can be measured. To date, more than 100 exoplanets have been detected around more than 1000 evolved stars \citep{jofre15,niedzielski15}. In one of these efforts, \cite{johnson10a} analyzed a sample of 246 subgiants from the SPOCS IV catalog, as well as 948 FGKM dwarfs and subgiants from the Keck M Dwarf Survey \citep{butler06} and the original SPOCS catalog \citep{vf05}. Using the complete sample of 1194 stars, with masses ranging from 0.2~\msun to 2.0~\msun, \cite{johnson10a} determined that the occurrence rate of giant planets increases approximately linearly with stellar mass, from $\sim$3\% for M dwarfs, to $\sim$8\% for FGK dwarfs, to $\sim$14\% for stars more massive than 1.5~\msun, the so-called "Retired A stars." \textbf{This correlation appears to be supported, at least for the FGKM dwarfs, by results from transit surveys \citep{fressin13,gaidos13}.} \cite{johnson10a} also confirmed that the planet-metallicity correlation holds for both M dwarfs and the Retired A Stars. This suggests that planet formation is significantly affected by the stellar mass and chemical composition---two  stellar properties that provide crucial links to the physical properties of stars' protoplanetary disks.

\textbf{These findings were questioned by \cite{lloyd11}, who pointed out that some discrepancies between the rotational velocities distributions of evolved planet hosts and field stars could be traced back to erroneous mass determinations. Moreover, Lloyd found that the numbers of massive subgiants (\mstar $\gtrsim$ 1.5 \msun) observed by \cite{johnson10a} seems inconsistent with the predictions from Galactic stellar population models.} More massive stars are formed less frequently according to the initial mass function (IMF) and also have a relatively rapid evolution across the Hertzprung Gap. Based on these arguments, Lloyd concludes that many of the Retired A Stars should in fact lie in the mass range 1.0--1.2 \msun. The errors on the original masses could stem from systematic
uncertainties on spectroscopically determined atmospheric parameters or ambiguities in the predictions from stellar
evolution models. 

\cite{johnson13} addressed the second concern by generating Galactic stellar population models with TRILEGAL \citep{girardi05}. Using a magnitude-limited sample (instead of the volume-limited sample used by \citealt{lloyd11}), they produced a simulated distribution that allows for a larger number of massive stars than Lloyd found---an effect similar to the Malmquist bias---and is thus consistent with the masses determined by \cite{johnson10a}. \cite{lloyd13}, on the other hand, argues that the mass distribution is not sensitive to the Malmquist bias, but instead to the usage of different Galactic models or input parameters within a given model. Using the Besan\c{c}on model \citep{robin03}, \citet{lloyd13} obtains a mass distribution with less massive subgiants that can not be reconciled with the observed data from \cite{johnson10a}.

The fraction of massive subgiants within the Retired A Star sample was also questioned by  \cite{sw13}. Based on an analysis that is independent of stellar evolution models, they showed that the space velocity dispersion of the subgiants with planets were larger than those of a sample of main-sequence A5--F0 stars, but consistent with the that obtained for F5--G5 dwarfs. They thus conclude that the subgiants with planets are most likely the evolved counterparts of less massive main-sequence solar-type stars, instead of ``Retired A Stars''. Their analysis also revealed that no more than 40\% of the planet-hosting subgiants could have been A5--F0 stars more massive than $\sim$1.3 \msun\ while on the main sequence.

In another study, \cite{sousa15b} pointed out that, while the masses they determined for planet-host stars were in general good agreement with values from the literature, there were some notable exceptions. In particular, differences of up to 100\% were found for a few evolved stars. Note, however, that some of these inconsistencies are explained by erroneous mass determinations in the original papers, as explained by \cite[][see Section 4.2]{tt15}. 

More recent studies of the stellar mass dependence of planet occurrence have been performed by \cite{reffert15} and \cite{jones16}. In the former study, the analysis of 373 G and K giants reveals that giant planet occurrence increases with stellar mass in the range 1.0 -- 1.9 \msun, consistent with the findings of \cite{johnson10a}. Moreover, there seems to be a maximum for the occurrence rate as a function of stellar mass at 1.9$^{+0.1}_{-0.5}$ \msun\ and a rapid decrease for more massive stars. The latter study uses a sample of 166 giant stars and finds a consistent result: occurrence rate increases with stellar mass, with the maximum at 2.1 \msun.

In the previous works mentioned, masses for the evolved stars were estimated by comparing observed properties (effective temperature, metallicity, and luminosity or surface gravity) with grids of stellar evolution models (e.g., \citealt{bressan12}). Although the accuracy of this method has been thoroughly tested for main-sequence stars (e.g., \citealt{torres10}), results for sugbiants and giants are still debated, as is clear from the above discussion. In an additional effort to test if stellar evolution models are also reliable for evolved stars, \cite{gj15} showed that PARSEC evolutionary tracks (\citealt{bressan12}) coupled with the code PARAM \citep{dasilva06} can recover model-independent masses from eclipsing binaries and asteroseismology within $\sim$4\% for the interval $\sim$0.7 -- 4.5 \msun. Therefore, the method itself does not present any issues that would significantly overestimate the masses of subgiants and giants. Nevertheless, incorrect input parameters could still lead to erroneous mass determinations. 

To avoid model-dependent issues, most recent efforts to solve the Retired A Stars controversy focused on the determination of empirical masses. \cite{johnson14} performed multiple mass measurements for the bright giant HD\,185351 using spectroscopic, interferometric and asteroseismic data. Although discrepancies on the 2.6$\sigma$ level were observed, all determined values confirm that HD 185351 is more massive than 1.5 \msun. \textbf{\cite{campante17} determined an asteroseismic mass for the planet-hosting subgiant HD\,212771 that is consistent with recent spectroscopic estimates and also with its classification as a Retired A Star.} \cite{stassun17} determined virtually model independent masses (from empirical radii and density or spectroscopic surface gravity) for 358 planet-hosting stars, achieving a precision better than 15\% for 134 of them. Within this most accurate sample, 30 stars lie in the HR diagram region typically occupied by the Retired A Stars and $\sim$80\% of them have masses consistent with this classification. 

\cite{north17} and \cite{stello17} investigated the masses of evolved stars using asteroseismology. The former study show that five of the seven analyzed objects (1.0 -- 1.7 \msun) have spectroscopic masses slightly larger than the corresponding asteroseismic value. However, the authors claim the offset is not significant and highly dependent on the adopted literature mass. In a similar comparison, the latter study reveals a 15-20\% offset for six (out of seven) stars with masses larger than 1.6 \msun. Both studies highlight a large scatter among literature masses and trace the differences back to the input parameters and their uncertainties used in the determination. 

In this work, we revisit the Retired A Stars sample with the goal of checking if their original masses were in fact overestimated, thereby an artificial correlation between this physical parameter and the occurrence rate of giant planets. We also take the opportunity to investigate the concerns related to the kinematics of these stars. The paper is organized as follows. The sample is presented in Section \ref{sample} and new spectroscopic and kinematical analyzes are described in Section \ref{analysis}. Our results are validated in Section \ref{validation}, including comparisons with the previous ones from \cite{johnson10a}, other literature sources and other methods. \textbf{In Section \ref{discussion}, we discuss the consistency between masses and space and rotational velocities and obtain an updated relation for the occurrence rate of giant planets as a function of stellar metallicity and mass.} Finally, our concluding remarks are presented in Section \ref{conclusions}.  

\section{Sample and Data}
\label{sample}

Our sample of subgiants consists of 245 stars that have been continuously monitored by the Lick and Keck subgiant planet surveys since 2004 and 2007, respectively (see Table \ref{table_sample}). The details regarding the sample selection are described in \cite{johnson06,johnson10b}. In summary, targets were selected from the \textit{Hipparcos} catalog \citep{esa97,vanleeuwen07}) according to the following criteria: 0.5 $< M_{V} <$ 3.5 and 0.55 $< B-V <$ 1.10 and $V \lesssim$ 8.5. Stars lying less than 1 mag above the main sequence defined by \cite{wright05} or in the clump ($B-V >$ 0.8 and M$_{V} <$ 2.0) region are excluded. 

High-resolution spectra for stars in both the Lick and Keck surveys were obtained with the HIRES (High Resolution Echelle Spectrometer; \citealt{vogt94}) spectrograph on the Keck I 10-m telescope (Mauna Kea, Hawaii). \textbf{Deckers B1, B2, B3, B5, C2 and E2 were used  with a binning of 3x1. This instrumental setup produces resolutions $R \simeq$ 50,000 -- 100,000\footnote{https://www2.keck.hawaii.edu/inst/hires/slitres.html} and an almost complete spectral coverage from $\sim$3600 \AA\, to $\sim$7990 \AA, except for an inter-detector gap from $\sim$6420 \AA\, to $\sim$6543 \AA\, and some inter-order spacings redward of 6600 \AA.} We use the ``template" spectra taken without the iodine cell. They were reduced with the Keck pipeline following standard procedures. We measured the signal-to-noise (S/N) values using 112 apparent continuum regions between 5220 \AA\, and 6860 \AA, carefully selected using the spectra of the Sun (reflected off Vesta) and HD 185351 as references. Our targets have S/N $\gtrsim$ 100 and the typical value is $\sim$220 (see Table \ref{table_sample}).

\section{Analysis and Results}
\label{analysis}

\subsection{The Line List}
\label{line_list}

We compiled the  initial line list for Fe I and Fe II from multiple sources: \cite{sousa08}, \cite{ghezzi10a}, \cite{schuler11}, \cite{tsantaki13}, \cite{sousa14}, \cite{liu14} and \cite{bedell14}. Only spectral features with $\lambda >$ 5000 \AA\ are selected because of line crowding for lower wavelengths. Lines in the intervals 6270 \AA\ $\leq \lambda \leq$ 6330 \AA\ and $\lambda >$ 6865 \AA\ are removed in order to avoid possible contamination by telluric lines. Using the HIRES solar spectrum as a reference, we selected only those Fe I lines that were relatively isolated and unblended. We were more flexible for the Fe II case due to the more limited number of lines available for this species. The HIRES solar spectrum was also used as a reference to exclude lines located in the inter-detector or inter-order gaps. 

We retrieved all atomic parameters (wavelength $\lambda$, excitation potential $\chi$, log\,\textit{gf} and van der Waals damping factor) from the Vienna Atomic Line Database (VALD; \citealt{ryabchikova15} and references therein) on January 2016.  We measured equivalent widths (EWs) for all lines on the Solar Flux Atlas \citep{kurucz84} with the updated version of ARES \citep{sousa15a}. We adopted the following parameters: \textit{smoothder} = 4, \textit{space} = 3.0, \textit{rejt} = 0.999, \textit{lineresol} = 0.1 and \textit{miniline} = 5. A few lines presented clearly wrong EWs and these were replaced by manual measurements done using the task \texttt{splot} in IRAF\footnote{Image Reduction and Analysis Facility (IRAF) is distributed by the National Optical Astronomy Observatory (NOAO), which is operated by the Association of Universities for Research in Astronomy, Inc. (AURA) under cooperative agreement with the National Science Foundation (NSF).}.

Using the above list, a Kurucz ATLAS9 ODFNEW model atmosphere \citep{ck04} for the Sun (effective temperature \teff = 5777 K, surface gravity \logg = 4.44, metallicity \feh\footnote{\feh = $\log(N_{Fe}/N_{H})_{\star}$ - $\log(N_{Fe}/N_{H})_{\odot}$} = 0.00 and microturbulence $\xi$ = 1.00 km s$^{-1}$) and the \textbf{driver \textit{abfind} of the} July 2014 version of the Local Thermodynamic Equilibrium (LTE) line analysis code MOOG\footnote{http://www.as.utexas.edu/~chris/moog.html} \citep{sneden73}, we derived individual Fe abundances for all lines. \textbf{Option 1 was used for the treatment of the damping in MOOG, i.e. tabulated van der Waals damping factors \citep{barklem00,baj05} are used when available within the code and values from VALD (see Table \ref{table_line_list}) are adopted otherwise.} We removed the Fe I lines that returned absolute abundances A(Fe)\footnote{A(Fe) = $\log(N_{Fe}/N_{H}) + 12$} lower than 7.20 or higher than 7.80. A similar cut was not applied to Fe II due to the more restricted number of lines for this species. We also excluded lines with reduced equivalent widths $\log$(RW) $\equiv$ $\log$(EW/$\lambda$) lower than -6.0 and larger than -4.8 in order to avoid too weak or saturated lines, respectively. Finally, we derived solar $\log$ \textit{gf} values by imposing that all remaining lines returned A(Fe) = 7.50 \citep{asplund09} after running MOOG with the same model atmosphere as above. \textbf{They were used in the analysis of our sample (see Section \ref{atm_par}) in order to improve the precision of the results, since laboratory and theoretical $\log$ \textit{gf} values for many iron lines still suffer from large uncertainties \citep[e.g.,][]{mashonkina11}.}

The final line list contains 158 Fe I and 18 Fe II lines (see Table \ref{table_line_list}). Using the the updated version of ARES \citep{sousa15a}, we automatically measured equivalent widths for the lines in this final list for all 245 target stars. The parameters were the same as those used for the Solar Atlas, except for \textit{rejt}, which was chosen according to the S/N value of each individual spectrum. 

\subsection{Atmospheric Parameters}
\label{atm_par}

We determined a homogeneous set of atmospheric parameters (\teff, \logg, \feh\ and \vmic) for our target stars using the standard spectroscopic method, which is based on the excitation and ionization equilibria of Fe I and Fe II. As for the Solar Atlas, the analysis was performed in LTE using the July 2014 version of MOOG with option 1 for the damping parameter. We used the ATLAS9 ODFNEW grid from \cite{ck04} to obtain interpolated model atmospheres.

We obtained the final parameters for each star through an automated iterative process that has to simultaneously meet four criteria: zero slopes in the linear fits between A(Fe I) and $\chi$ (excitation equilibrium) and between A(Fe I) and $\log$(RW); same average values of A(Fe I) and A(Fe II) (ionization equilibrium); and same value for the metallicity in the input model atmosphere and the output result from MOOG. The convergence in each criterion is achieved by successive changes in \teff, \vmic, \logg\ and \feh, respectively. Note that \feh\ = $\langle$A(Fe I)$_{\star}\rangle$ - A(Fe)$_{\odot}$, where we adopted A(Fe)$_{\odot}$ = 7.50 \citep{asplund09}. The iterative process also included two rounds of 2$\sigma$ clipping in order to remove lines that returned abundances too discrepant from the average values.

We calculated uncertainties in the atmospheric parameters in the following way. The microturbulence was changed until the linear fit between A(Fe I) and $\log$(RW) had a slope equal to the error of the zero slope from the converged solution. The difference between the this new \vmic\ and the best value is adopted as the uncertainty in this parameter. A similar procedure was done for the effective temperature, but the linear fit between A(Fe I) and $\chi$ was used instead. \textbf{Moreover, the contribution from the error in \vmic\ is taken into account by varying this parameter by its uncertainty determined in the previous step and checking what is the slope produced in the linear fit between A(Fe I) and $\chi$. Then, \teff is varied until we recover a zero slope and the difference between this temperature and best one is taken as the contribution from the error in \vmic. The uncertainty in the surface gravity is obtained by varying this parameter until the difference between A(Fe I) and A(Fe II) is equal to the standard deviation of the mean for the latter abundance, taken from the converged solution. The contribution of \teff\ to the error on \logg\ was also considered (in a procedure similar to the one used above to estimate contribution of \vmic\ to the uncertainty on \teff).} Finally, the uncertainty in \feh\ takes into account the standard deviation of mean for A(Fe I) and the variations caused by the errors in \teff, \vmic\ and \logg, all added in quadrature.

As a first test for our method, we analyzed the HIRES spectrum of sunlight reflected off Vesta (S/N = 267) as if it was a regular target star. The only difference is that only one round of 2$\sigma$ clipping was performed. Otherwise, many good lines would be excluded due to the low value of $\sigma$ (as expected, since the $\log$ \textit{gf} values were adjusted using the Sun as a reference). The determined atmospheric parameters are: \teff\ = 5778 $\pm$ 20 K, \logg\ = 4.45 $\pm$ 0.07, \feh\ = 0.01 $\pm$ 0.01 and \vmic\ = 1.009 $\pm$ 0.030 km s$^{-1}$. They are all in excellent agreement with the canonical solar values. 

\begin{figure}
\plotone{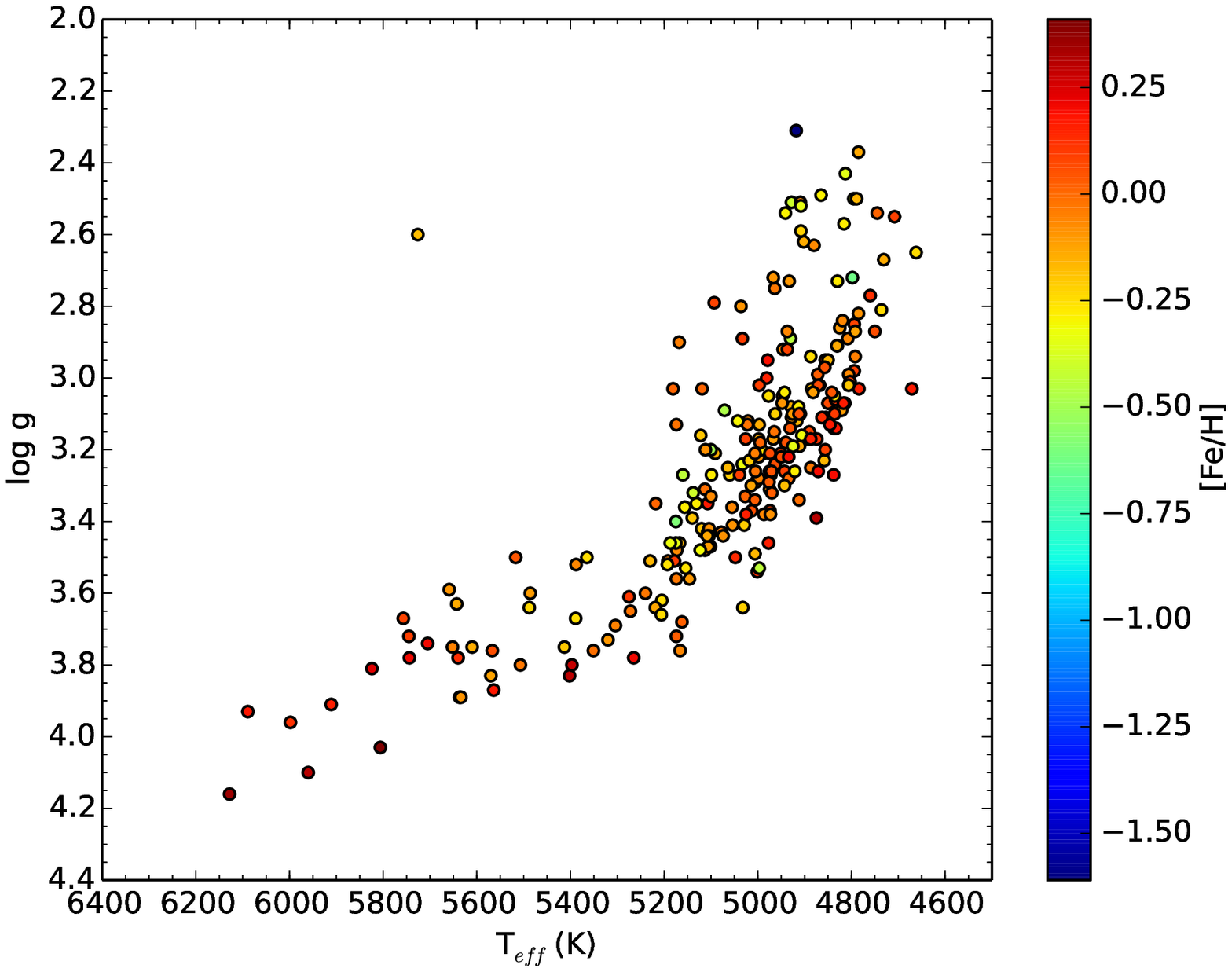}
\caption{Atmospheric parameters (\teff, \logg\ and \feh) for our sample of 245 subgiants.}
\label{plot_atm_par}
\end{figure}

The final atmospheric parameters for our target stars are shown in Figure \ref{plot_atm_par}. The individual values as well as the associated uncertainties can be seen in Table \ref{table_atm_par}. We should note that the relatively low errors presented on this table are the internal uncertainties of the spectroscopic analysis. \textbf{More realistic external uncertainties for the standard spectroscopic method employed here (used in the discussion of Section \ref{offsets_atm_par}) should be higher, with typical values of up to $\sim$100 K for \teff, $\sim$0.2 dex for \logg, $\sim$0.2 dex for \feh\ and $\sim$0.2 km s$^{-1}$ for \vmic\ \citep[e.g.,][]{ghezzi10a,hinkel16}. For completeness, we show in Table \ref{atm_par_sensitivities} the sensitivities of the atmospheric parameters to these larger uncertainties for HD 185351 (first spectrum), taken as a representative star in our sample.}

The three results for HD 185351 come from the independent analysis of three different spectra obtained during one observing night and reveal an excellent consistency between the derived parameters. For Figure \ref{plot_atm_par} and the subsequent analyzes in this study, we adopted the arithmetic means as the final atmospheric parameters for HD 185351: \teff\ = 5039 $\pm$ 16 K, \logg\ = 3.27 $\pm$ 0.06, \feh\ = 0.09 $\pm$ 0.01 and \vmic\ = 1.110 $\pm$ 0.020 km s$^{-1}$. The uncertainties were calculated through simple error propagation. It is worth noting that the final \logg\ value is in excellent agreement with the asteroseismic value (3.280 $\pm$ 0.011) presented by \cite{johnson14}.

Two stars in Figure \ref{plot_atm_par} clearly differ from the rest of the sample. The first is HD 150331, which apparently has a \logg\ (2.60 $\pm$ 0.09) that is too low or a \teff\ (5726 $\pm$ 35 K) that is too high relative to the region occupied by our sample. The Geneva-Copenhagen (GC) Catalog \citep{casagrande11} reports an effective temperature similar to ours (5867 $\pm$ 93 K), but a discrepant surface gravity (3.35). SIMBAD Astronomical Database shows it as a G0/2 bright giant, a classification that is more consistent with our parameters. The GC metallicity (0.42) is also very different from ours (-0.18). If we use a model atmosphere with the GC parameters and a microturbulence velocity \vmic\ = 1.000 km s$^{-1}$ as an input to the driver \textit{ewfind} in MOOG, we do not obtain EWs consistent with those measured with ARES. 

The second star is HD 21581 and it is significantly more metal-poor ([Fe/H] = -1.61) than all other stars in the sample. Nevertheless, this apparent discrepancy is not an issue since its metallicity, as well as \teff\ and \logg\, are consistent with other literature measurements (see the 2016 version of the PASTEL catalogue; \citealt{soubiran16}). Thus, we believe our parameters for these two stars are reliable and use them in the following analyzes. 

\subsection{\textbf{Rotational Velocities}}
\label{vsini}

\textbf{We determined the stellar projected rotational velocities \vsini\ for all our stars through spectral synthesis with MOOG. The driver \textit{synth} was used with option 1 for the damping. The Fe I line at 6703.565 \AA\ was chosen for the analysis because it is relatively isolated and unblended. The atomic parameters adopted for the line are the same as in Table \ref{table_line_list}. Model atmospheres for each star are also necessary as input and we used those obtained in Section \ref{atm_par} for the final atmospheric parameters.} 

\textbf{The synthesis in MOOG was performed using the option \textit{r} for the broadening of the spectral line. This means that separate values for the instrumental profile FHWM$_{inst}$, the macroturbulence velocity $V_{macro}$ and the rotational velocity \vsini\ have to be provided. The values for FWHM$_{inst}$ were chosen according to the setup used for observing each star: 0.096 \AA\ for deckers B1, B2 and B3; 0.112 \AA\ for deckers B5 and C2; 0.065 \AA\ for decker E2. These values correspond to resolutions of $\sim$70,000, $\sim$60,000 and $\sim$100,000, respectively and can be seen in Table \ref{table_vsini}. We calculated the values of $V_{macro}$ (see Table \ref{table_vsini}) using the relations presented by \cite{brewer16}, using our values for \teff\ and \logg\ (the same as in the model atmospheres described above) in the equations. We should note that, although \feh\ does not appear on the relations, HD 21581 has a significantly lower metallicity than the stars used to build the calibrations.}

\textbf{Small adjustments to the continuum level (typically 0.5\%) and shifts to the central wavelength of the Fe line (typically 0.02 \AA) were allowed in order to account for uncertainties in the normalization process and to properly match the observed feature, respectively. We also let the A(Fe) abundance vary within 0.06 dex (with steps of 0.02 dex) the [Fe/H] value determined for the star to account for intrinsic line-to-line scatter in the abundances. After these values were set, we computed a grid of synthetic spectra for \vsini\ values varying between 0.0 and 15.0 km s$^{-1}$ with steps of 0.2 km s$^{-1}$. The best match between the observed and synthetic spectra were found through the minimization of the reduced $\chi^{2}$:}

\begin{equation}
\chi_{r}^{2} = \frac{1}{(d-1)}\sum_{i=1}^{n}\frac{(O_{i} - S_{i})^{2}}{\sigma^{2}}
\label{chi2red}
\end{equation}

\begin{figure}
\plotone{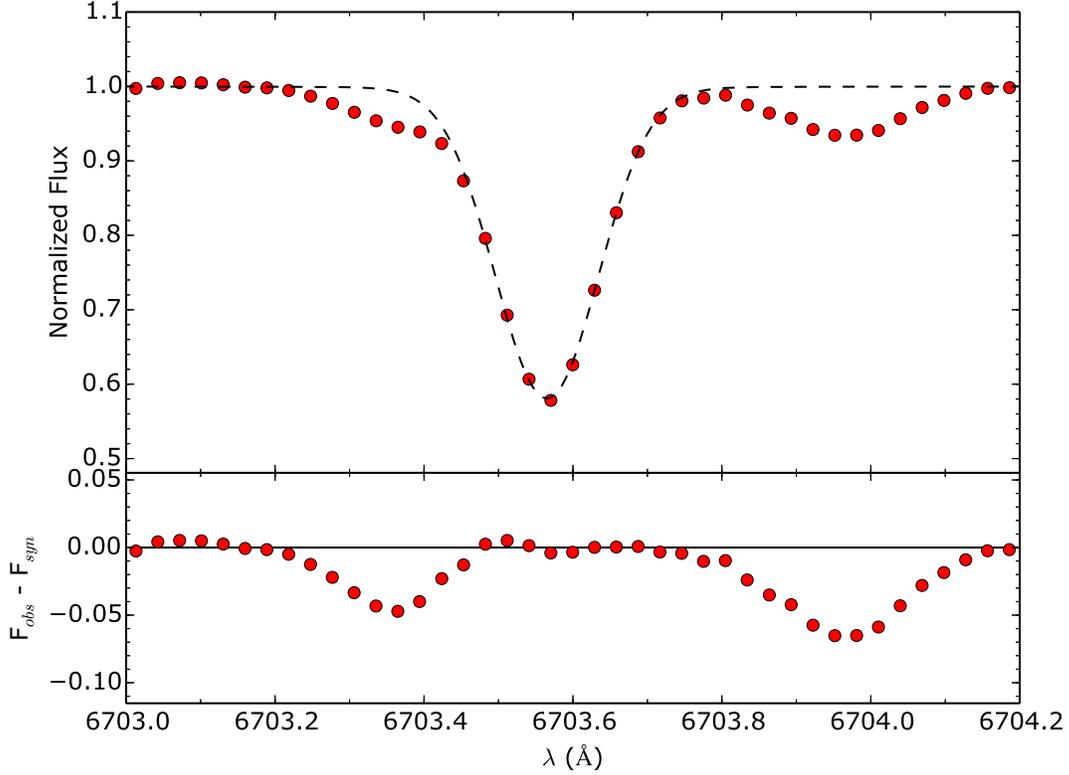}
\caption{\textit{Upper panel:} Observed (red circles) and best fit synthetic spectrum (black dashed line) for the second spectrum of HD 185351. Only the Fe I line at 6703.565 \AA\ was considered in the fit. \textit{Lower panel:} Differences between the observed and synthetic spectrum (red circles). The black solid line represents a null difference.}
\label{plot_vsini}
\end{figure}

\textbf{In the above equation, $O_{i}$ and $S_{i}$ correspond, respectively, to the observed and synthetic normalized fluxes at point \textit{i} of the spectral line profile. The rms of the continuum is given by $\sigma$ = (S/N)$^{-1}$ (see Table \ref{table_sample} for S/N values) and \textit{d = n - p} represents the number of degrees of freedom in the fit, where \textit{n} is the number of points in the observed spectra and \textit{p = 1} is the number of free parameters in the analysis (only \vsini). We calculated the uncertainties on \vsini\ by varying this parameter until $\Delta\chi_{r}^{2} = \chi_{r}^{2} - \chi_{r,min}^{2} = 1$. The values of the rotational velocities and its associated errors are shown in Table \ref{table_vsini}. As an example, we show the best fit for the second spectrum of HD 185351 in Figure \ref{plot_vsini}. For this star, we can see a very good consistency between the three sets of results. As for the atmospheric parameters, we decided to adopt the arithmetic mean and the propagated error as the final value \vsini\ = 1.7 $\pm$ 0.4 km s$^{-1}$.}

\textbf{As an initial test for our method, we once again analyzed the HIRES spectrum of sunlight reflected off Vesta (S/N = 267) as if it was a regular target star. Using FHWM$_{inst}$ = 0.096 \AA\ (decker B2) and the calculated $V_{macro}$ = 3.57 km s$^{-1}$, we obtained 1.4 $\pm$ 0.8 km s$^{-1}$, which is in very good agreement with the typical solar value ($\approx$1.6 km s$^{-1}$; e.g., \citealt{vf05,pavlenko12,brewer16}). In a more general test, we computed synthetic spectra for the Fe I 6703.565 \AA\ line using a model atmosphere calculated for the typical atmospheric parameters in our sample (\teff\ = 5050 K, \logg\ = 3.30, \feh\ = 0.00 and \vmic\ = 1.000 km s$^{-1}$), the corresponding $V_{macro}$ (3.45 km s$^{-1}$), FHWM$_{inst}$ = 0.112 \AA\ (the lower resolution) and \vsini values ranging from 0.0 to 10.0 km s$^{-1}$, with steps of 0.5 km s$^{-1}$. Then, we used the method described above to analyze these spectra as if they belonged to real stars and were able to recover all rotational velocities within 0.5 km s$^{-1}$.}

\subsection{Evolutionary Parameters}
\label{evol_par}

We determined the evolutionary parameters of our target stars (mass, radius, \logg\ and age) through the comparison of observational parameters with grids of evolutionary tracks. We performed this analysis using the version 1.3 of the PARAM code, kindly provided by Leo Girardi\footnote{For the web interface maintained by Leo Girardi at the Osservatorio Astronomico di Padova, visit http://stev.oapd.inaf.it/cgi-bin/param.}. The code adopts a Bayesian statistical framework in order to determine masses, radii, surface gravities and ages from given sets of priors and input parameters \citep{dasilva06}. \textbf{We used version 1.1 of the PARSEC evolutionary tracks \citep{bressan12} as well as the default options for the Bayesian priors: lognormal Initial Mass Function (IMF) from \cite{chabrier01}, constant Stellar Formation Rate (SFR), and the age interval from 0.1 to 12.0 Gyr. We also chose to provide the input parameters \teff, \feh, $V$ magnitude and parallax $\pi$ because asteroseismic information is not available for most of our targets.}

The effective temperatures and metallicities were homogeneously and spectroscopically determined on this study (see Section \ref{atm_par}). The adopted parallaxes and associated uncertainties (see Table \ref{table_sample}) are the revised values from \cite{vanleeuwen07}. The only exception was the star HD 33298, which had the negative parallax replaced by the original value from the \textit{The Hipparcos and Tycho Catalogues} \citep{esa97}. We calculated the $V$ magnitudes from the formula $V = V_{T}-[0.090\times(B_{T}-V_{T})]$, with values of B$_{T}$ and V$_{T}$ taken from the \textit{Tycho-2} catalog \citep{hog00}. Uncertainties were determined through simple error propagation. 

We corrected all $V$ magnitudes for extinction using $A_{V}$ values calculated with an adapted version of the code \texttt{extinct.for} \citep{hakkila97}, which uses the tables from \cite{arenou92} (see Table \ref{table_sample}). \textbf{These tables remain among the most relaible sources for estimating reddening for nearby stars within 280 pc \citep[e.g.,][]{gontcharov17}. Uncertainties on the corrected magnitudes V$_{0}$ were obtained through simple error propagation, where $A_{V}$ uncertainties were estimated using the relative errors $\sigma_{A_{V}}/A_{V}$ given in the tables from \cite{arenou92}.} 

\begin{figure}
\plotone{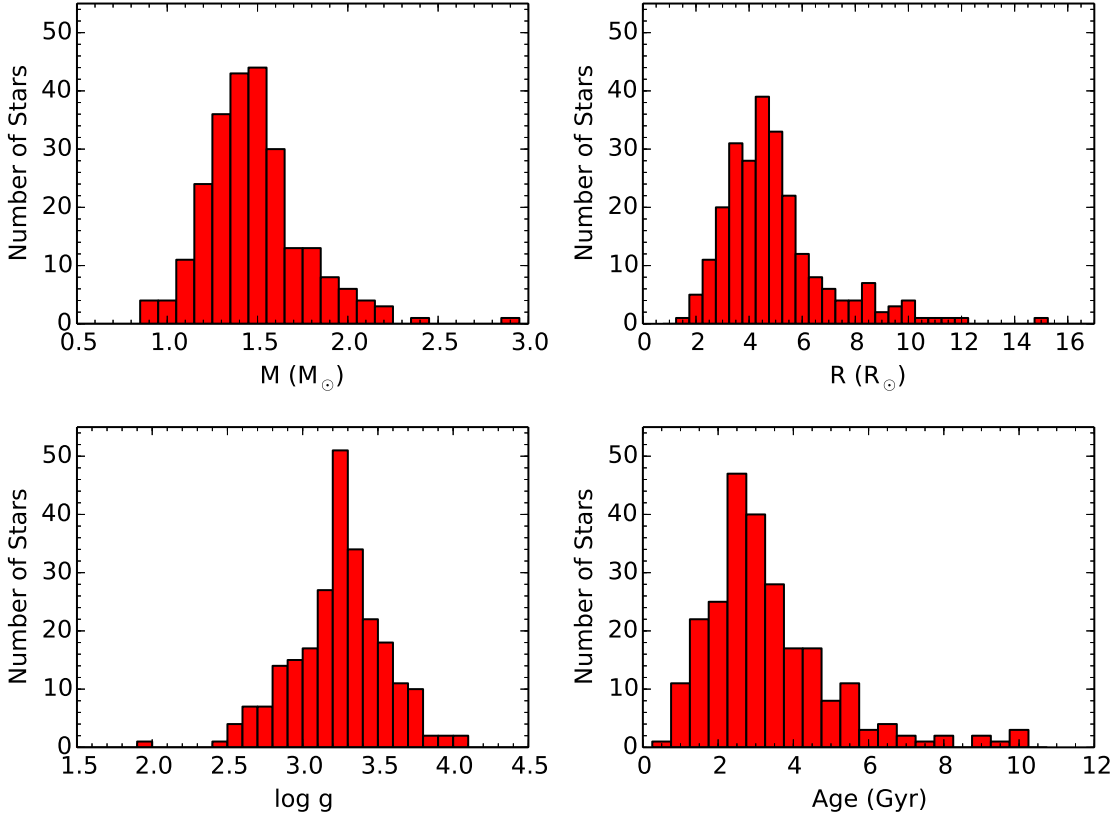}
\caption{Distributions for the evolutionary parameters for our sample of 245 subgiants. Masses, radii, surface gravities and ages are shown in the upper left, upper right, lower left and lower right panels, respectively.}
\label{plot_evol_par}
\end{figure}

The distributions for the evolutionary parameters are shown in Figure \ref{plot_evol_par}. The individual values for each star are presented on Table \ref{table_evol_par} along with their uncertainties. The histograms confirm that our stars are mostly more massive, evolved and younger than the Sun, as expected from the target selection. It should also be noted that the mass distribution is consistent with the simulation presented in Figure 3 of \cite{johnson13}. 

\subsection{Kinematics}
\label{kinematics}

\begin{figure}
\plotone{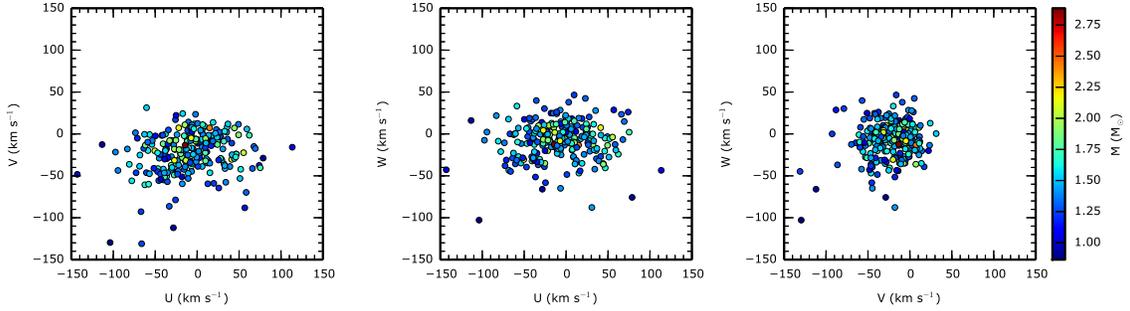}
\caption{\textbf{Galactic \textit{UVW} space velocities for 244 of the subgiants (HD 10425 did not have an available radial velocity). Masses are represented by the color scale.}}
\label{plot_kinematics}
\end{figure}

We calculated galactic \textit{UVW} space velocities for our target stars using an adapted version of the code \texttt{gal\_uvw.py}\footnote{Originally written by W. Landsman and made available by Sergey Koposov on astrolibpy at GitHub.}. Equatorial coordinates (right ascencion $\alpha$ and declination $\delta$), parallaxes and proper motions were taken from \cite{vanleeuwen07} (see Table \ref{table_sample}). \textbf{The California Planet Survey uses the iodine cell technique of \cite{butler96} to compute highly precise RVs relative to each star's own spectrum, and therefore produces relative RV measurements. We use absolute RVs, where available, for our kinematics study from the following references, in order of preference: \cite{chubak12}, \cite{af12}, \cite{gontcharov06}, \cite{massarotti08} and \cite{brewer16} (see Table \ref{table_sample}). The star HD 10245 did not have a RV available in any of these references and therefore its space velocities were not calculated.} The resulting \textit{UVW} velocities can be seen on Table \ref{table_kinematics} and their distributions are presented in Figure \ref{plot_kinematics}. Note that $U$, $V$ and $W$ are positive towards the Galactic center, in the direction of Galactic rotation and in the direction towards the North Galactic pole, respectively. We have not applied corrections for the solar motion in order to compare our results to those presented by \cite{sw13}. 

\section{Validation of the Results}
\label{validation}

As discussed in Section 1, possibly overestimated masses in \cite{johnson10a} could have been produced by systematic uncertainties on the input atmospheric parameters. Although the results derived in our study are homogeneous and precise, their accuracy has yet to be tested. We do this by comparing our spectroscopic parameters with those derived by independent methods or studies.

\subsection{Validation of the Spectroscopic Temperatures}
\label{test_spec_teff}

As a first check, we calculated photometric effective temperatures for our sample using the \teff\ versus $V-K_{S}$ calibrations from \cite{gh09}. We calculated the $V$ magnitudes from B$_{T}$ and V$_{T}$ taken from the \textit{Tycho-2} catalog \citep{hog00}, as explained in Section \ref{evol_par}. We adopted $K_{S}$ magnitudes from the Two Micron All Sky Survey (2MASS) All-Sky Catalog of Point Sources \citep{cutri2003} (see Table \ref{table_phot_teff}). We removed 15 stars with flags D, E or F for the $K_{S}$ magnitude. Moreover, we could not find a reasonable match for the star HD 136418 (i.e., with a $K_{S}$ value that would provide a reasonable \teff) within the 2MASS catalog. 

We corrected the $V-K_{S}$ colors for the reddening in the following way. \textbf{First, we converted $A_{V}$ values derived in Section \ref{evol_par} to $E(B-V)$ using the formula $A_{V} = 3.1 \times E(B-V)$.} Then, we calculated $E(V-K_{S})$ values from $E(B-V)$ using the relation $E(V-K_{S}) = 2.70 \times E(B-V)$, where the coefficient was taken from \cite{rm05}. The final dereddened color $(V-K_{S})_{0}$ was obtained with the formula $(V-K_{S})_{0} = (V-K_{S}) - E(V-K_{S})$. 

The calculated sets of photometric effective temperatures and its errors are presented in Table \ref{table_phot_teff}. It should be noted that \cite{gh09} derive different polynomials for dwarfs/subgiants and giants, with the limit placed at \logg\ = 3.00 (Gonz\'alez Hern\'andez, private communication). We followed this separation when choosing which calibration would be applied to each star, using the spectroscopic \logg\ values as the reference. We calculated the uncertainties in the photometric effective temperatures by adding two contributions in quadrature: the errors propagated from $(V-K_{S})_{0}$ and \feh\ and the standard deviation of the final calibration (32 K for the dwarfs/subgiants and 23 K for the giants). 

\begin{figure}
\plotone{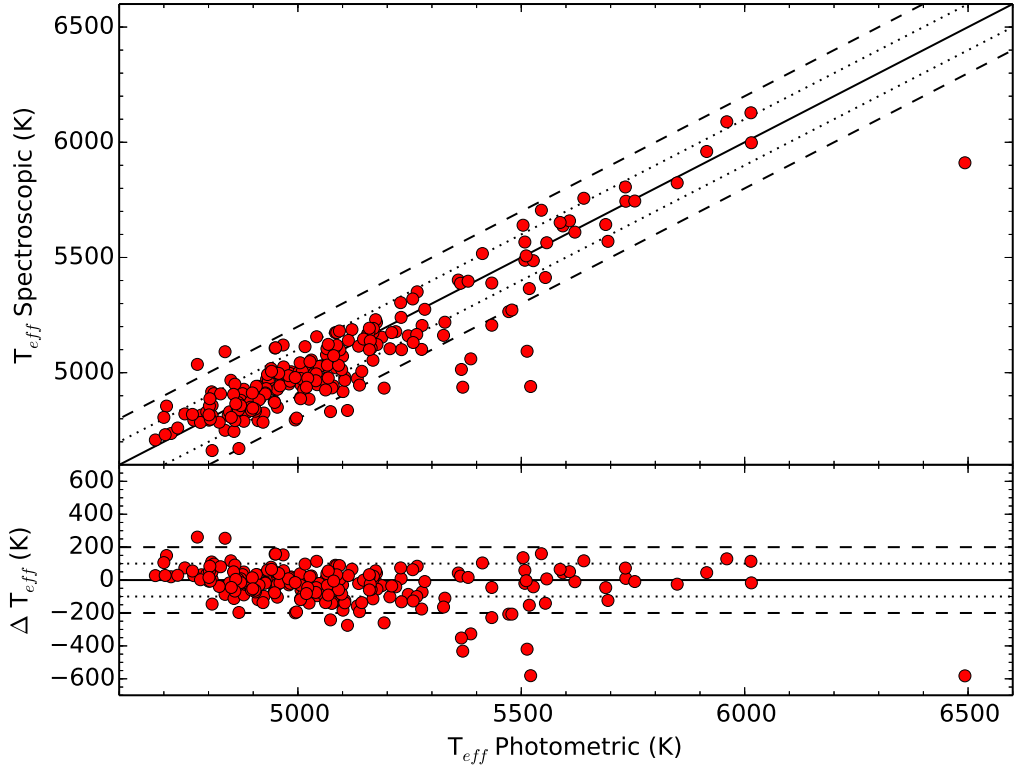}
\caption{Comparison between spectroscopic and photometric effective temperatures for 229 out of 245 target stars (see text for the reasons why some stars were removed). The upper panel shows the direct comparison between the two sets of temperatures. The lower panel presents the difference $\Delta$\teff = $T_{eff}^{Spec}$ - $T_{eff}^{Phot}$ as a function of the photometric \teff. In both panels, solid lines represent a perfect agreement, while dotted and dashed lines correspond to $\pm$100 K and $\pm$200 K, respectively.}
\label{plot_phot_teff}
\end{figure}

The photometric temperatures are compared with the spectroscopic values in Figure \ref{plot_phot_teff}. We can observe a general good agreement over the whole \teff\ interval. Most stars present offsets lower than $\pm$200 K, and the average difference (in the sense spectroscopic - photometric) is -27 $\pm$ 111 K. If we do not consider the six stars with differences larger than $\pm$300 K, the mean difference is -16 $\pm$ 87 K. These values validate our spectroscopic temperature scale and are consistent with the typical external uncertainty ($\sim$100 K) mentioned in Section \ref{atm_par}. Moreover, both the differences and the scatters are similar to those obtained by \cite{gh09} when their results are compared with spectroscopic temperatures (see their Section 8.4). 

\subsection{Validation of the Spectroscopic Gravities}
\label{test_spec_logg}

\begin{figure}
\plotone{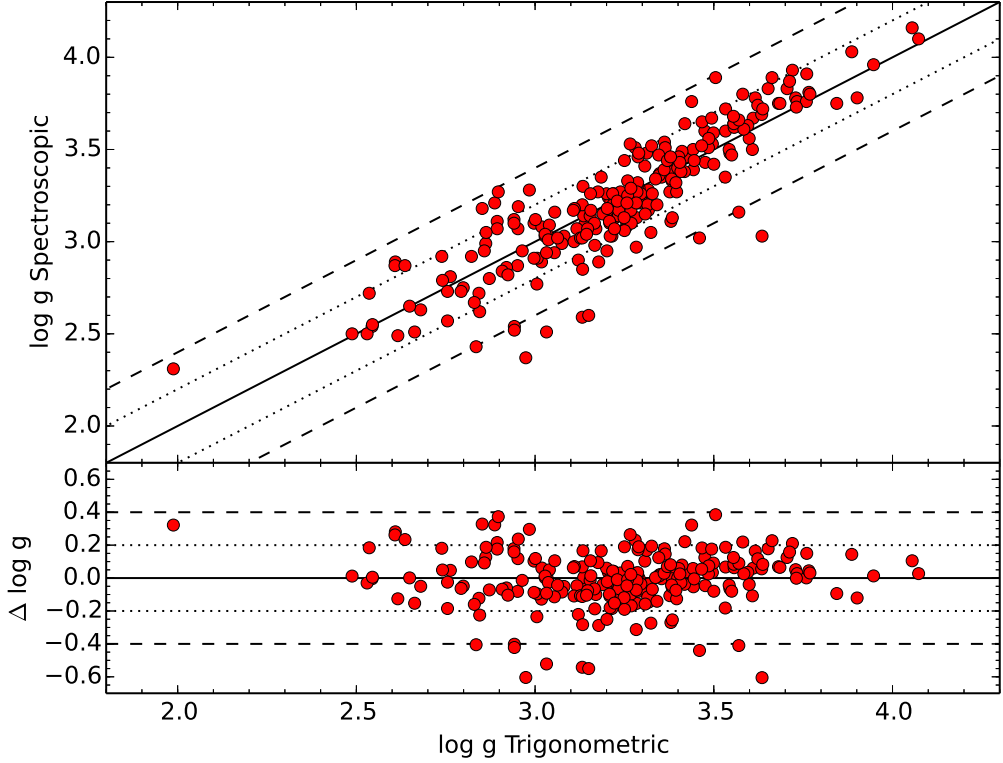}
\caption{Comparison between spectroscopic and trigonometric surface gravities for our 245 target stars. The upper panel shows the direct comparison between the two sets of gravities. The lower panel presents the difference $\Delta$\logg = \logg$^{Spec}$ - \logg$^{Trig}$ as a function of the trigonometric \logg. In both panels, solid lines represent a perfect agreement, while dotted and dashed lines correspond to $\pm$0.2 dex and $\pm$0.4 dex, respectively.}
\label{plot_comp_logg}
\end{figure}

The second test consisted in the comparison of the spectroscopic surface gravities with the trigonometric values returned by PARAM in Section \ref{evol_par} (see Table \ref{table_evol_par}). \textbf{We can see in Figure \ref{plot_comp_logg} that the global agreement is good and most stars have offsets smaller than $\pm$0.4 dex, considered as a 2$\sigma$ limit, where $\sigma$ is taken as, e.g., the value of the external uncertainty for \logg\ quoted in Section \ref{atm_par} or the systematic offset between spectroscopic and trigonometric surface gravities discussed by \cite{dasilva06}.} The mean difference (in the sense spectroscopic - trigonometric) is -0.008 $\pm$ 0.163 dex. If the ten stars with differences larger than $\pm$0.4 are not taken into account, the average difference becomes 0.012 $\pm$ 0.131 dex. We should note that the outliers in this case are note the same as the ones found in the comparison of the temperatures. \textbf{As for \teff, the above differences are consistent with the typical external uncertainty ($\sim$0.2 dex) mentioned in Section \ref{atm_par}.}

\begin{figure}
\plotone{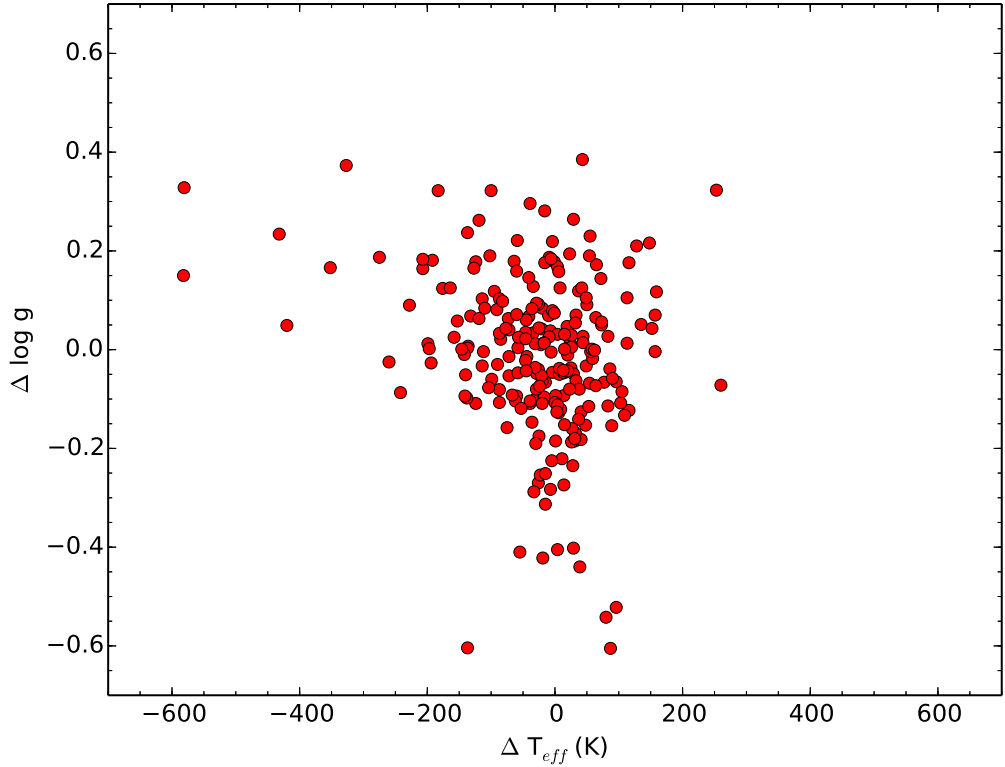}
\caption{Differences between spectroscopic and trigonometric surface gravities as a function of the differences between spectroscopic and photometric effective temperatures.}
\label{plot_diff_teff_diff_logg}
\end{figure}

We also checked if there was any correlation between the differences in \teff\ and \logg\ because systematically incorrect effective temperatures could cause systematic offsets in the surface gravities in the spectroscopic method. We can see there is no such correlation in Figure \ref{plot_diff_teff_diff_logg}. A linear fit to the data returns a correlation coefficient $R^{2}$ = 0.06. Another evidence that our surface gravities are reliable is the excellent agreement between the spectroscopic values obtained for the star HD 185351 and the asteroseismic measurement presented by \cite{johnson14}, as discussed in Section \ref{atm_par}. 

\subsection{Validation of the Metallicities}
\label{test_spec_feh}

Photometric metallicities are not as precise as the spectroscopic ones, so they do not provide a good reference to validate our results. In order to do that, we decided to perform an analysis of the Hyades stars studied by \cite{df16}. The reduced spectra were kindly made available by Let\'icia Dutra-Ferreira and the S/N values were taken from Table 1 of the corresponding paper. We analyzed all 17 stars (14 dwarfs and 3 giants) in the same way as our targets (see Section \ref{atm_par}). 

The average metallicity found for the cluster was $\langle$\feh$\rangle$ = 0.16 $\pm$ 0.04, which is in very good agreement with the value recommended by \cite{df16} (0.18 $\pm$ 0.03). The mean metallicities for the dwarfs (0.16 $\pm$ 0.04) and the giants (0.15 $\pm$ 0.02) are in excellent agreement, which shows that our line list is capable of providing a consistent metallicity scale for stars in a wide range of effective temperatures and in different evolutionary stages. \textbf{Moreover, we do not find any trends in \feh\ as a function of \teff, which is an additional evidence that our metallicity scale is consistent.} Finally, the average metallicity obtained for the giants is consistent with the value obtained by \cite{df16} (0.14 $\pm$ 0.03) when using an optimized line list in a classical spectroscopic analysis. This result supports the fact that the metallicity scale for the subgiants is reliable.

\subsection{Comparison with the SPOCS Catalogues}
\label{test_spocs}

\begin{figure}
\plotone{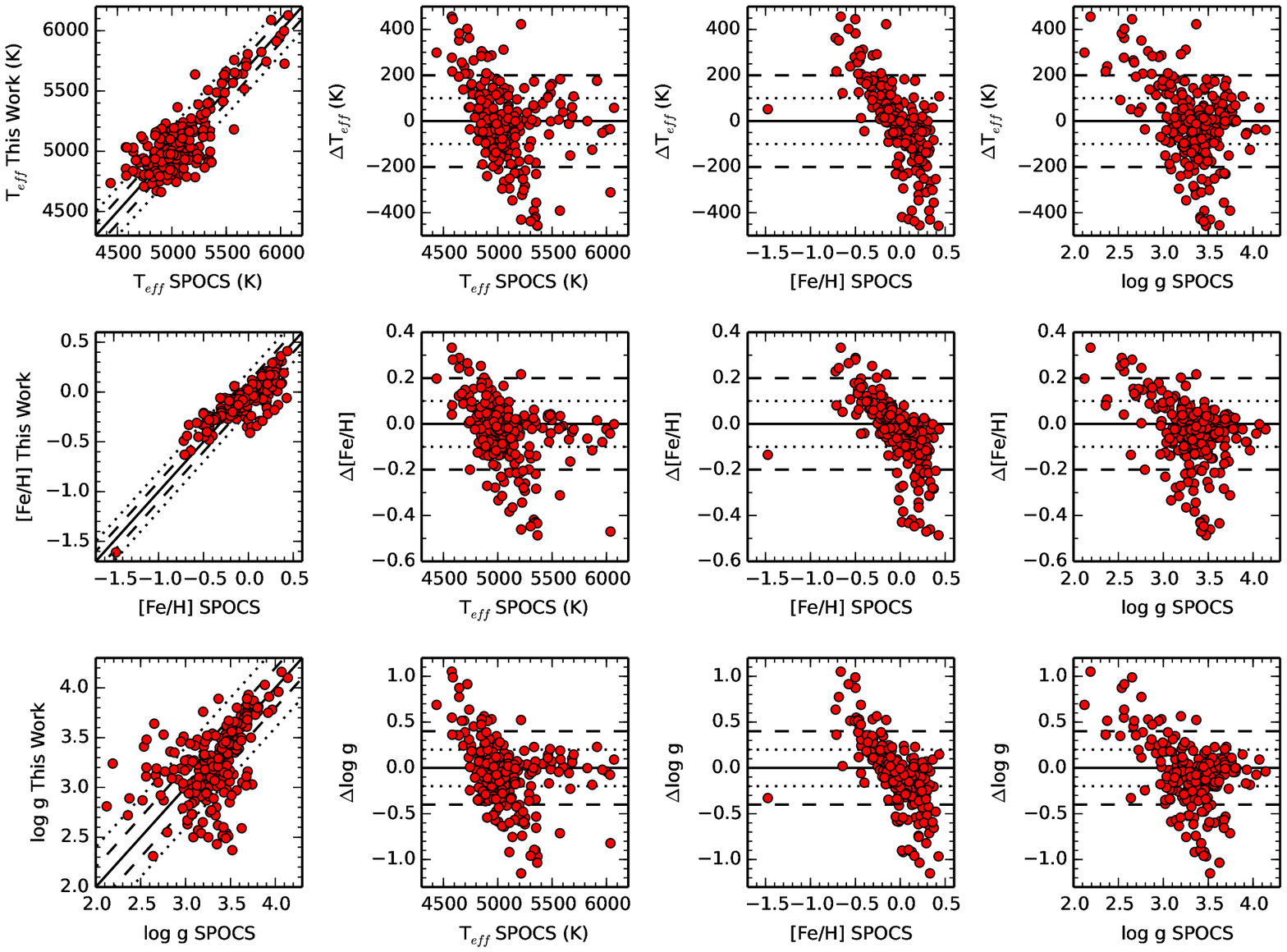}
\caption{Comparison between the atmospheric parameters derived in this work (see Section \ref{atm_par}) and those used by \cite{johnson10a}. The left column of panels shows the direct comparisons between the two sets of parameters. The other three columns present the differences between the two sets of parameters (in the sense this work - SPOCS) as a function of, from left to right, \teff, \feh\ and \logg\ from the SPOCS catalog. In all panels, solid lines represent a perfect agreement. Dotted and dashed lines correspond, respectively, to $\Delta$\teff = $\pm$100 K and $\pm$200 K, $\Delta$\feh = $\pm$0.1 dex and $\pm$0.2 dex or $\Delta$\logg = $\pm$0.2 dex and $\pm$0.4 dex.}
\label{plot_comp_spocs}
\end{figure}

The parameters used by \cite{johnson10a} were obtained from the Spectroscopic Properties of Cool Stars (SPOCS) catalog and derived through spectral synthesis using the Spectroscopy Made Easy (SME) software \citep{vf05}. They thus provide an additional independent check for the results derived in this work. All stars in our sample have SME parameters and the average differences (in the sense this work - SPOCS) are: $\Delta$\teff\ = -8 $\pm$ 165 K, $\Delta$\feh\ = -0.04 $\pm$ 0.14 dex and $\Delta$\logg\ = -0.08 $\pm$ 0.34 dex. Although the mean differences are reasonable, the dispersions are larger than the typical external uncertainties. This is also clear from Figure \ref{plot_comp_spocs}, where we can also see some trends in the residuals, especially as a function of \feh\ ($R^{2}$ = 0.50, 0.41 and 0.42 for $\Delta$\teff, $\Delta$\feh\ and $\Delta$\logg, respectively).

\begin{figure}
\plotone{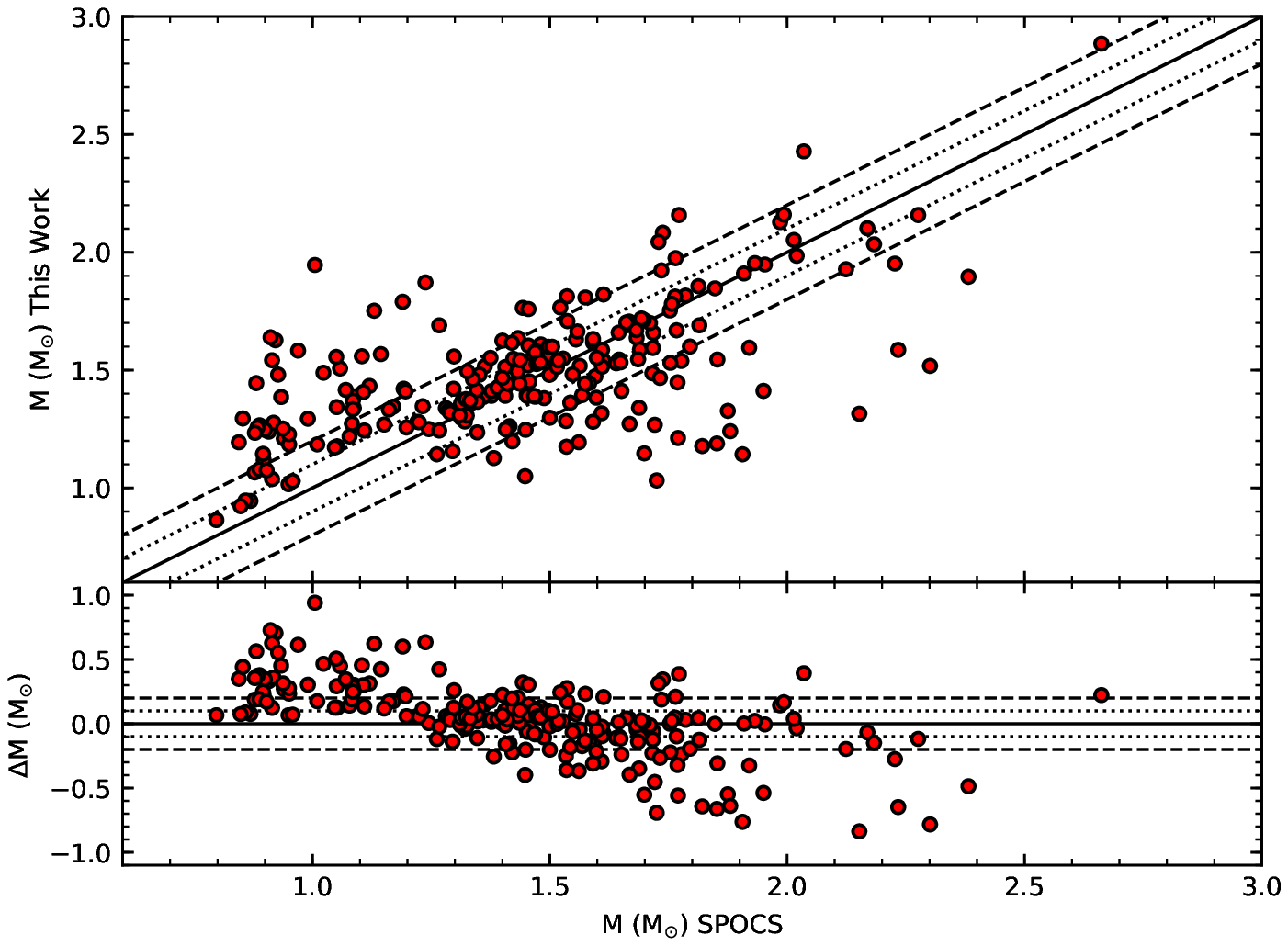}
\caption{Comparison between the masses derived in this work (see Section \ref{evol_par}) and those used by \cite{johnson10a}. The upper panel shows the direct comparison between the two sets of parameters. The lower panel presents the difference $\Delta$\mstar = \mstar (This Work) - \mstar (SPOCS) as a function of the latter. In both panels, solid lines represent a perfect agreement, while dotted and dashed lines correspond to $\pm$0.1 \msun\ and $\pm$0.2 \msun, respectively.}
\label{plot_comp_mass_spocs}
\end{figure}

A similar comparison for the masses yields an average difference (in the sense this work - SPOCS) of $\Delta$\mstar = 0.05 $\pm$ 0.28 \msun. Again, the mean difference is good, but the dispersion too large. As clearly seen in Figure \ref{plot_comp_mass_spocs}, a significant trend is present in the residuals ($R^{2}$ = 0.43), probably a consequence of the trends observed for the atmospheric parameters. 

The most recent version of the SPOCS catalog \citep{brewer16} contains atmospheric parameters and chemical abundances for 1626 FGK stars. For the 204 stars in common with our data set, the average differences (in the sense this work - SPOCS) are: $\Delta$\teff\ = 48 $\pm$ 38 K, $\Delta$\feh\ = -0.04 $\pm$ 0.05 dex and $\Delta$\logg\ = -0.01 $\pm$ 0.09 dex. Note that \cite{brewer16} adopted a lower reference solar iron abundance (A(Fe) = 7.45), so we applied a correction of -0.05 dex for all SPOCS [Fe/H] values in order to put their metallicities into our scale. There is a considerable improvement in the dispersions, however a few trends in the residuals are still present, in particular for $\Delta$\feh\ versus \logg\ ($R^{2}$ = 0.58), getting stronger for the more evolved stars (\logg\ $\lesssim$ 3.5). 

Similar trends are found by \cite{brewer16} when comparing their results with other spectroscopic analyzes, but their origin remain unclear. Possible explanations could be the strong correlations observed between \teff, \feh\ and \logg\ when spectral synthesis is used \citep[e.g.,][]{torres12} or the fixed value for the microturbulence (0.85 km s$^{-1}$) adopted in SME. We stress that the standard spectroscopic analysis employed in this work shows no such correlations \citep{torres12} and our results are in good agreement with those obtained from independent methods.

\begin{figure}
\plotone{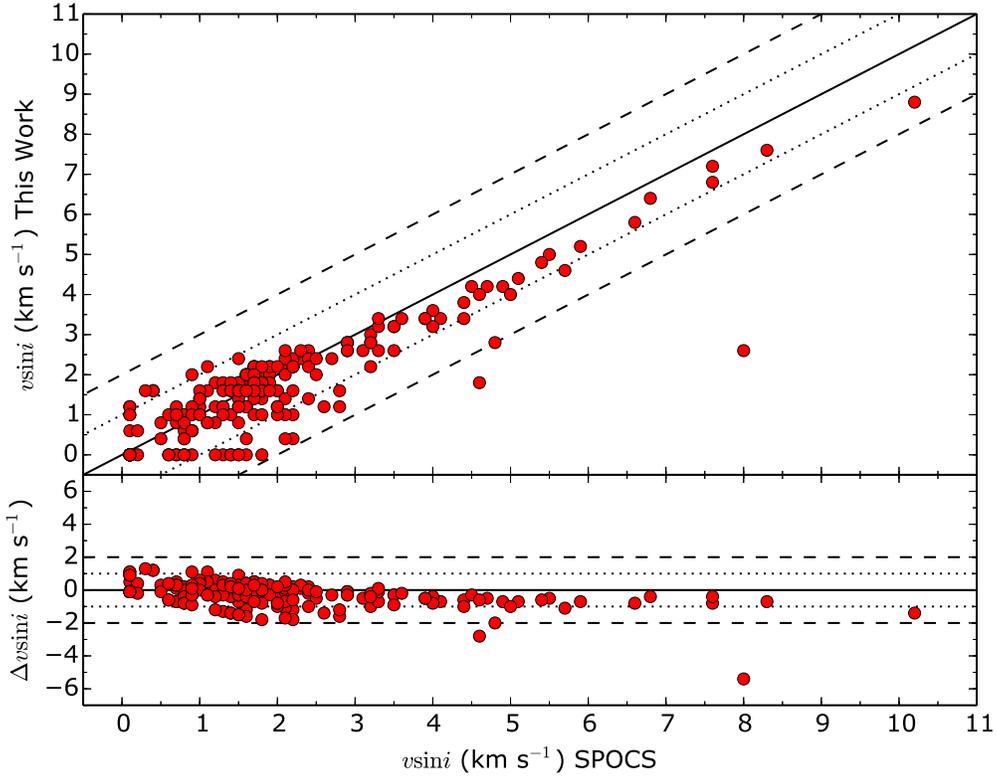}
\caption{Comparison between the rotational velocities derived in this work (see Section \ref{vsini}) and those determined by \cite{brewer16}. The upper panel shows the direct comparison between the two sets of parameters. The lower panel presents the difference $\Delta$\vsini\ = \vsini\ (This Work) - \vsini\ (SPOCS) as a function of the latter. In both panels, solid lines represent a perfect agreement, while dotted and dashed lines correspond to $\pm$1.0 km s$^{-1}$ and $\pm$2.0 km s$^{-1}$, respectively.}
\label{plot_comp_vsini}
\end{figure}

\textbf{\cite{brewer16} also determined rotational velocities and we compare our values with theirs in Figure \ref{plot_comp_vsini}. For the 204 stars in common, the average difference (in the sense this work - SPOCS) is $\Delta$\vsini\ = -0.25 $\pm$ 0.73 km s$^{-1}$. The stars with differences larger than 2.0  km s$^{-1}$ are HD 31543 (-5.4 km s$^{-1}$), a possible single-lined spectroscopic binary (SB1; \citealt{chubak12}), and HD 103616 (-2.8 km s$^{-1}$), a suspected binary \citep{nordstrom04}.}

\subsection{Possible Systematic Offsets on the Masses}
\label{test_mass}

\subsubsection{Possible Offsets in the Atmospheric Parameters}
\label{offsets_atm_par}

\textbf{In the previous sections, we showed that the spectroscopic atmospheric parameters of the subgiants are accurate. \cite{gj15} showed that PARSEC evolutionary tracks \citep{dasilva06} and the PARAM code are able to provide reliable masses for a sample of 59 benchmark subgiants and giants. Nevertheless, we decided to test if the masses of subgiants could be significantly overestimated if systematic offsets were present in the input atmospheric parameters derived in this work (\teff\ and \feh). In order to do this, we decreased all our effective temperatures and metallicities by 100 K and 0.10 dex, respectively. The offset in \teff\ was chosen so that we would have a decrease in the stellar masses and is consistent with typical external uncertainties (see Section \ref{atm_par}) found in the literature when different studies are compared (but higher than the internal errors derived here). We should also note that the offsets are consistent, in the sense that such a decrease of $\sim$100 K in the \teff\ for a typical Retired A Star (\teff\ $\sim$ 5000 K) would cause a decrease in the iron abundance of about $\sim$0.1 dex (see Table \ref{atm_par_sensitivities}).}   

\begin{figure}
\plotone{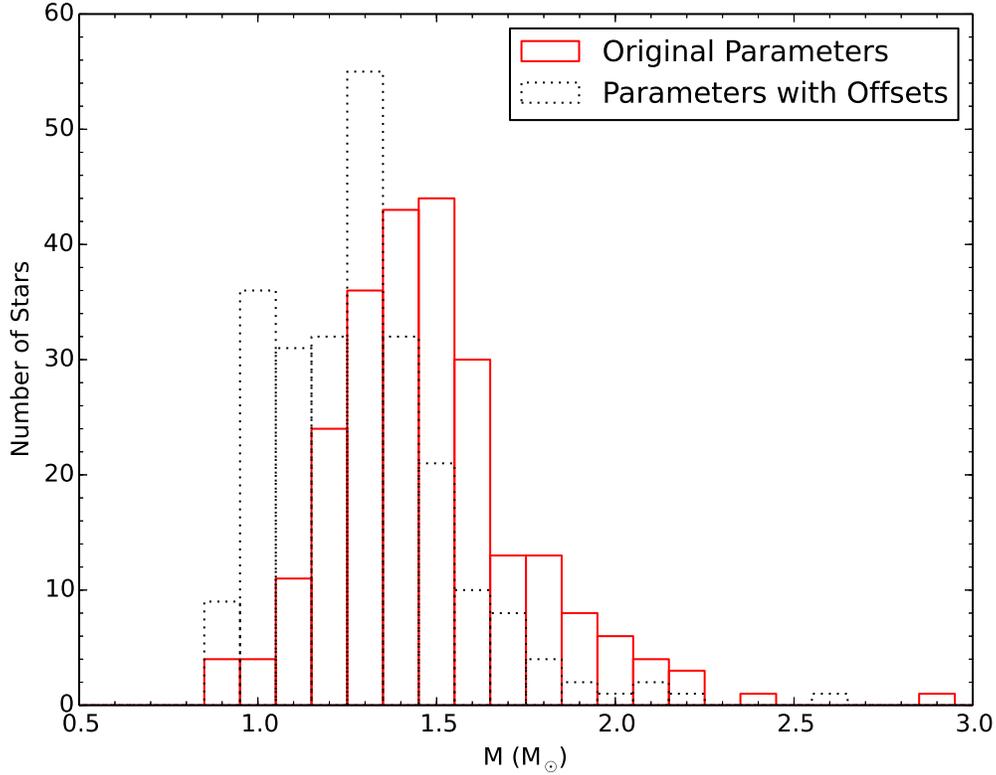}
\caption{Comparison between the mass distributions obtained with the original atmospheric parameters derived in this work (solid red histogram) and the ones obtained with the offsets of -100 K in \teff\ and -0.1 dex in \feh\ (black dotted histogram).}
\label{plot_mass_offset}
\end{figure}

Using these new parameters, we repeated the analysis done in Section \ref{evol_par}. The comparison of the new set of masses with the original ones can be seen in Figure \ref{plot_mass_offset}. As expected, the new masses are smaller than the original ones. The average difference (in the sense no offset - with offset in the atmospheric parameters) is 0.20 $\pm$ 0.14 \msun. Considering the typical uncertainty of the original masses is 0.11 \msun, this difference is consistent with a 2$\sigma$ offset and much smaller than the possible overestimation of up to 50\% suggested by \cite{lloyd11}. We should also highlight that the \teff\ and \feh\ offsets used in this test are much larger than the average difference between our spectroscopic and photometric temperatures (-27 K) and the offsets in the metallicities scales obtained for the Hyades stars (-0.02, with our values being smaller), respectively. Therefore, we feel confident that possible offsets in the atmospheric parameters are small (if present) and do not significantly overestimate the masses of the subgiants.

\subsubsection{Effect of the Reddening}
\label{test_reddening}

\begin{figure}
\plotone{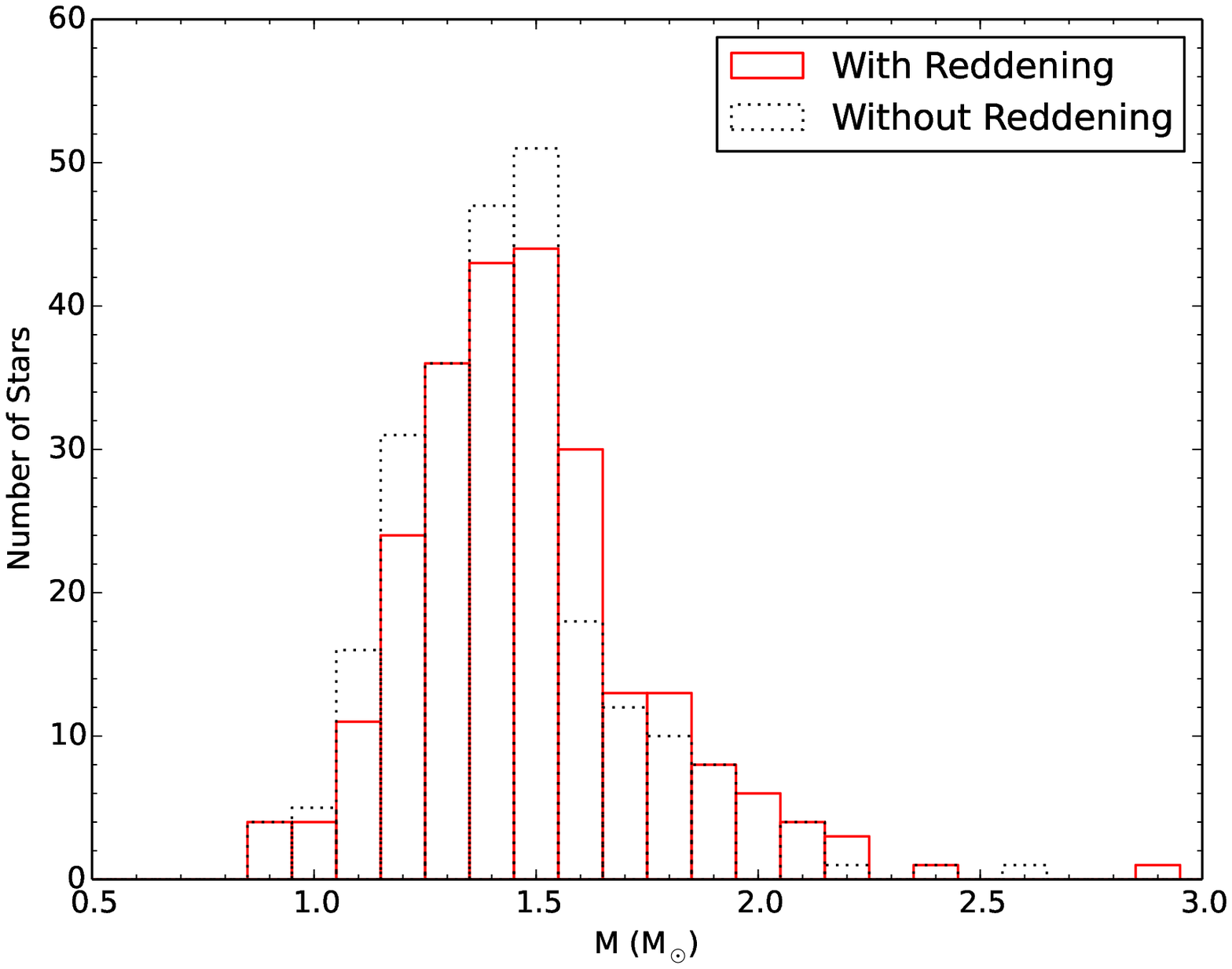}
\caption{Comparison between the mass distributions obtained with (solid red histogram) and without (black dotted histogram) correcting the $V$ magnitudes for the reddening.}
\label{plot_reddening}
\end{figure}

Another input parameter that can affect the determination of the masses is the reddening. In order to test how much the reddening corrections could overestimate our masses, we decide to repeat the analysis done in Section \ref{evol_par} considering $A_{V}$ = 0 for all stars in our sample. The comparison of these new results with the original ones is shown in Figure \ref{plot_reddening}. As expected, masses obtained without correcting the $V$ magnitudes for the reddening are smaller, however the difference is not significant. The mean difference (in the sense with reddening - without reddening) is 0.05 $\pm$ 0.05 \msun, a value that is smaller than the typical uncertainty of 0.11 \msun. Therefore, we see that the reddening correction does not significantly increase the masses of the subgiants.

\subsubsection{\textbf{Gaia's Parallaxes}}
\label{test_gaia}

\begin{figure}
\plotone{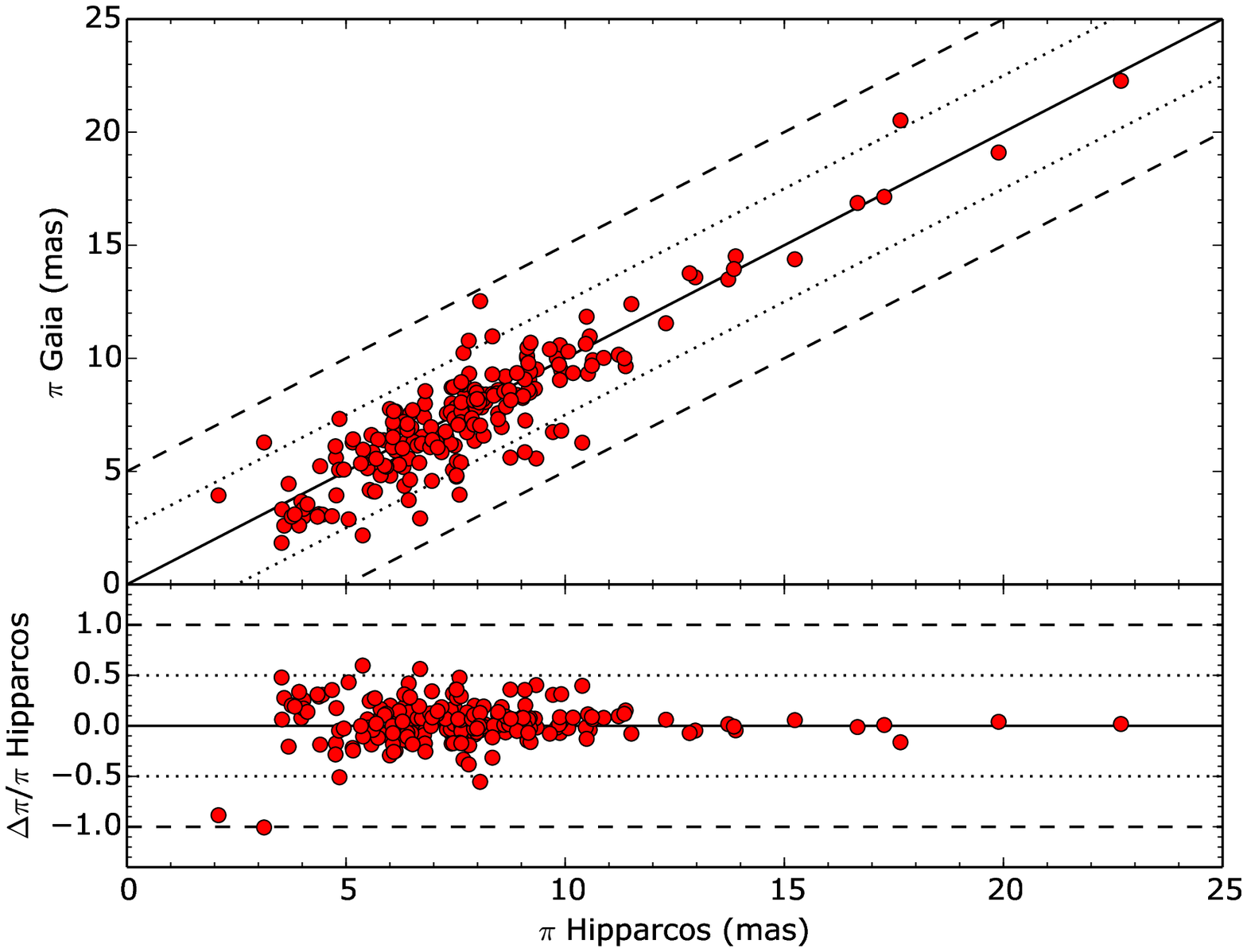}
\caption{\textbf{\textit{Upper panel:} Comparison between Gaia's new parallaxes and the Hipparcos revised values adopted in this work. The black solid line represents a perfect agreement, while dotted and dashed lines show $\pm$2.5 mas and $\pm$5 mas offsets, respectively. \textit{Lower panel:} Percentual differences (in the sense Hipparcos - Gaia) between the two sets of parallaxes. The black solid line shows a null difference. Dotted and dashed lines represent 50\% and 100\% differences.}}
\label{plot_comp_par}
\end{figure}

\textbf{Gaia's first data release\footnote{http://www.cosmos.esa.int/web/gaia/dr1} (DR1; Gaia Collaboration 2016) presents trigonometric parallaxes for $\sim$2 million stars in common with \textit{Hipparcos} \citep{esa97} and \textit{Tycho-2} Catalogues \citep{hog00}. We found 208 stars from our sample in DR1 and decided to compare Gaia's new parallaxes with the \textit{Hipparcos} revised values adopted in this work \citep{vanleeuwen07} (see Figure \ref{plot_comp_par}). The average difference (in the sense Hipparcos - Gaia) is 0.25 $\pm$ 1.30 mas or 3 $\pm$ 20 \%. We should highlight that \cite{st16} found the exact same systematic offset for a sample of eclipsing binaries with accurate empirical distances.}

\textbf{Although the average difference is small, there are a few stars with discrepancies between 50\% and 100\%. Thus, we decided to test if the masses of the subgiants are affected by the parallax choice. When using Gaia's parallaxes for the 208 stars in common (and not considering the remaining 37 stars), we obtain masses that are on average higher than those derived with \textit{Hipparcos} parallaxes by 0.08 $\pm$ 0.21 \msun\ or 6 $\pm$ 14 \%. We see that the choice of \textit{Hipparcos} parallaxes provides slightly lower masses, so they are not causing any systematic effects that would overestimate the masses of the subgiants. We decided not to use Gaia's parallaxes for now because the astrometric results from DR1 are still preliminary. However, we anticipate that this particular choice does not affect the conclusions drawn from this study.}

\subsection{Validation of Space Velocities}
\label{test_uvw}

\begin{figure}
\plotone{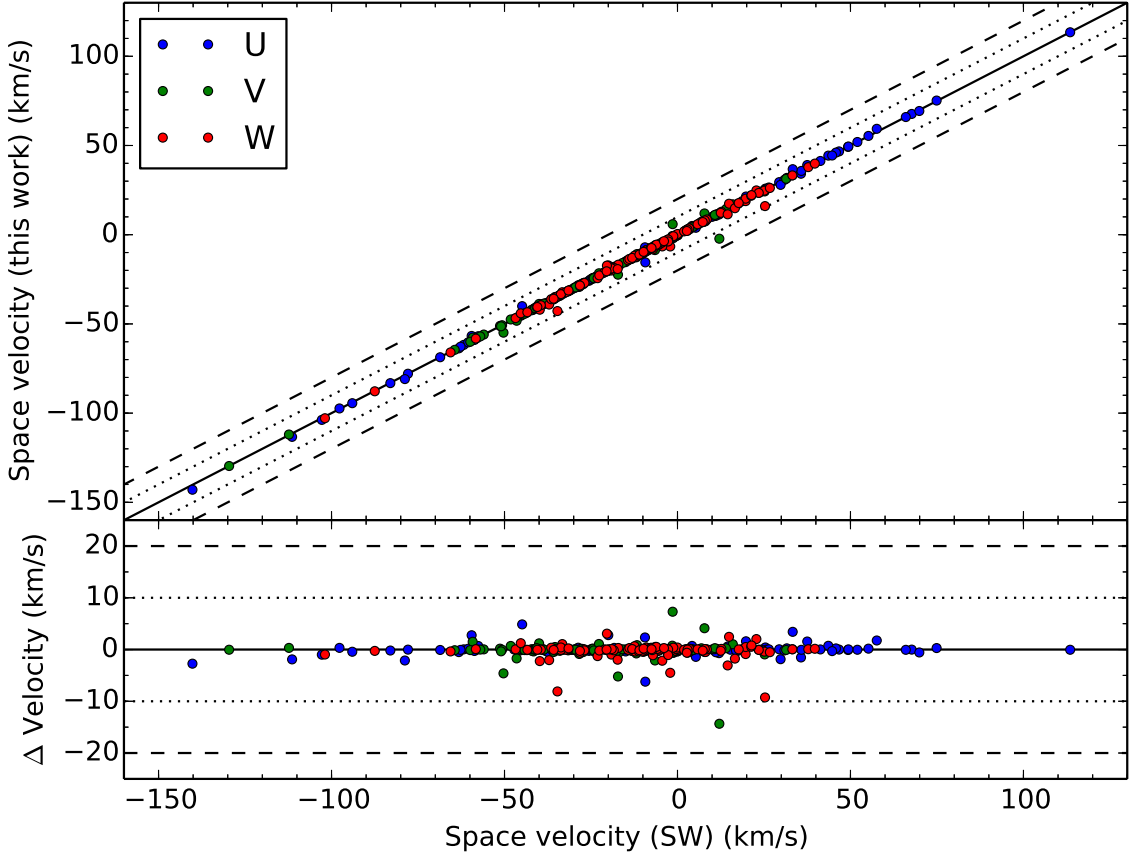}
\caption{Comparison between $UVW$ velocities (blue, green and red points, respectively) derived in this work and those from \cite{sw13}. The black solid line represents a perfect agreement. Dotted and dashed lines show, respectively, $\pm$10 km s$^{-1}$ and $\pm$20 km s$^{-1}$ offsets, arbitrarily chosen for reference. \textit{Lower panel:} Absolute differences (in the sense this work - SW) between the two sets of space velocities. The lines are the same as in the upper panel.}
\label{plot_comp_uvw_sw}
\end{figure}

Space velocities calculated in Section \ref{kinematics} can also be compared with values from the literature. Our sample has 114 stars that were also analyzed by \cite{sw13}. As can be seen in Figure \ref{plot_comp_uvw_sw}, there is a very good agreement between the two sets, with the following average differences (in the sense this work - SW): $\Delta$U = -0.01 $\pm$ 1.08 km s$^{-1}$, $\Delta$V = -0.15 $\pm$ 1.73 km s$^{-1}$ and $\Delta$W = -0.28 $\pm$ 1.42 km s$^{-1}$. Six stars (HD 11970, HD 126991, HD 158038, HD 180053, HD 193342 and HD 203471) present higher than average differences, which are related to different values adopted for the radial velocities. 

Our sample also has 28 stars in common with the Geneva Copenhagen Catalog \citep{casagrande11}. The agreement between the two sets of values is good and the average differences (in the sense this work - GC) are: $\Delta$U = 0.95 $\pm$ 2.81 km s$^{-1}$, $\Delta$V = 1.39 $\pm$ 6.17 km s$^{-1}$ and $\Delta$W = -0.97 $\pm$ 4.98 km s$^{-1}$. Three stars (HD 11970, HD 31543 and HD 64730) present large differences in the space velocities and are basically responsible for the relatively high dispersions around the averages. These differences can be traced back to distinct values adopted for the radial velocities. These two comparisons confirm that our $UVW$ velocities are consistent with previous values from the literature, and thus are reliable for being used in the following discussions.

\section{Discussion}
\label{discussion}

\subsection{Consistency between Masses and Kinematics}
\label{mass_uvw}

\textbf{One of the main concerns regarding the Retired A Stars is that their masses may not be consistent with their space velocities \citep{sw13}. If these stars are indeed the evolved counterparts of A dwarfs, we would expect their velocity dispersions to be smaller than those that would be observed if they were actually F and G stars while on the main sequence. Using solar neighborhood samples defined within the \textit{Hipparcos} catalog, \cite{sw13} showed that subgiant planet-hosting stars\footnote{Their sample is different from the ones analyzed here and in \cite{johnson10a}.} are more likely to have evolved from main-sequence F5-G5 rather than A5-F0 stars.} 

\begin{figure}
\plotone{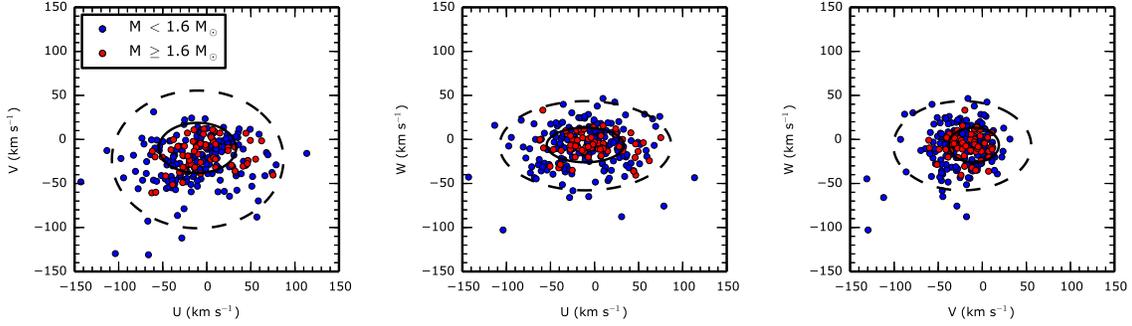}
\caption{\textbf{Galactic \textit{UVW} space velocities for 244 of our subgiants (HD 10245 did not have an available radial velocity).} Stars with masses \mstar $<$ 1.6 M$_{\odot}$ and \mstar $\geq$ 1.6 M$_{\odot}$ are represented by blue and red points, respectively. The black solid and dashed lines show the 95\% velocity ellipsoids for, respectively, the samples of main-sequence A5-F0 and F5-G5 \textit{Hipparcos} stars, as defined in \cite{sw13}.}
\label{plot_kinematics_2}
\end{figure}

\textbf{In Figure \ref{plot_kinematics_2}, we compare the space velocities derived in this work for the subgiants with the 95\% velocity ellipsoids for the samples of main-sequence A5-F0 and F5-G5 \textit{Hipparcos} stars, taken from \cite{sw13}. \textbf{It should be noted that these stars have mean distances of 100 $\pm$ 44 pc and 62 $\pm$ 23 pc, thus sampling the solar neighborhood. With an average distance of 141 $\pm$ 54 pc, our targets sample a very similar volume.} If we consider our targets with fractional uncertainties in the parallax lower than 20\% (following \citealt{sw13}), the fractions of subgiants within the A5-F0 95\% velocity ellipsoids are: 54\% for UV, 48\% for UW and 47\% for VW. However, if we restrict our sample to \mstar $\geq$ 1.6 M$_{\odot}$, these fractions increase to 67\%, 63\% and 71\%, respectively.}

\textbf{Following equation 1 from \cite{sw13}, we calculated $\delta_{i}$ values for each star in their A5-F0 and F5-G5 samples, as well as for our sample of subgiants. We compare these distributions using a Kolmogorov-Smirnov test and again considering only stars with fractional uncertainties in the parallax lower than 20\%. The probability that the subgiants and the A5-F0 stars are drawn from the same parent population is $\sim$10$^{-38}$, but much higher ($\approx$0.06) when compared to the F5-G5 stars. If we now use only Retired A Stars (i.e., stars with \mstar $\geq$ 1.6 M$_{\odot}$), the former probability increases to $\sim$10$^{-6}$ and the latter decreases to $\approx$0.05. Although the agreement is much better, this result still supports the idea that the kinematics of the Retired A Stars would imply they were in fact FG dwarfs while on the main sequence.}

\textbf{\cite{sw13} apparently did not use any reddening corrections before selecting the sample of A5-F0 stars. Using a similar procedure as the one described in Section \ref{evol_par}, we derived $A_{V}$ and $E(B-V) \approx A_{V}/3.1$ values for all stars in their sample. Applying these corrections, respectively, to their absolute magnitudes $M_{V}$ and colors $B-V$, we obtain a new group of main-sequence A5-F0 stars following the selection criteria in \cite{sw13}: 0.15 $< B-V <$ 0.30, 1.9 $< M_{V} <$ 2.7 and $\sigma_{\pi}/\pi <$ 0.20. Defining the new corresponding 95\% velocity ellipsoids, the fractions of Retired A Stars $\sigma_{\pi}/\pi <$ 0.20 within them are: 77\% for UV, 79\% for UW and 85\% for VW. These numbers are significantly higher than those obtained for the original ellipsoids from \cite{sw13}. Because reddening was not accounted for, the $M_{V}$ and $B-V$ intervals were including brighter and hotter stars, which naturally define smaller ellipsoids.} 

\textbf{As this effect could explain the apparent inconsistency between kinematics and mass, we applied reddening corrections to the selection of A5-F0 and F5-G5 stars and performed new KS tests. The probabilities that these samples and the Retired A Stars are drawn from the same parent population are $\approx$0.0003 and $\approx$0.02, respectively. Although reddening corrections make the A5-F0 sample more consistent with the Retired A Stars, it is not able to explain the apparent inconsistency between their kinematics and masses.}

\textbf{The last test was to check if the uncertainties on the masses contributed to this inconsistency. Using steps of 0.01 M$_{\odot}$ for offsets in the masses of the Retired A Stars, we discovered that, for a decrease of only 0.12 M$_{\odot}$, we can no longer reject the null hypothesis that they are drawn from the same parent population as the A5-F0 stars (p $\approx$ 0.03), as defined by \cite{sw13}. At this point, the probability that our stars and their F5-G5 sample are drawn from the same distribution is p $\approx$ 0.009. If reddening corrections are considered in the selection of the A5-F0 and F5-G5 samples (see discussion above), the transition occurs for an even smaller decrease in mass: 0.04 M$_{\odot}$. For this offset, the probabilities that the Retired A Stars and A5-F0 and F5-G5 stars are drawn from the same parent population are p $\approx$ 0.02 and p $\approx$ 0.008, respectively. Therefore, we can conclude that masses derived from evolutionary tracks and the kinematics of the Retired A Stars are consistent within the expected uncertainties of these two different approaches, especially if reddening is considered.}    

\subsection{\textbf{Consistency between Masses and Rotational Velocities}}
\label{mass_vsini}

\textbf{Another concern regarding the Retired A Stars is that their masses may not be consistent with their rotational velocities \citep{lloyd11}. Stellar rotation is a strong function of both mass and evolutionary stage. Stars born with masses above the Kraft break ($\sim$1.3\msun; \citealt{kraft67}) have progressively thinner convective envelopes, or no surface convection at all. Therefore, angular momentum loss through magnetic winds is negligible or absent during the main sequence and they remain as rapid rotators until the end of this evolutionary stage. As these stars evolve across the subgiant branch however, a convective zone develops in the expanding and cooling envelope and magnetic winds are also produced. These effects combined decrease the stellar rotation rate. Therefore, the analysis of the rotation of subgiants provides a powerful diagnostics for stellar mass \citep[e.g.,][]{vsp13}.}

\textbf{\cite{lloyd11} noted that three of the evolved planet hosts with 1.6 \msun\ $<$ \mstar $<$ 2.0 \msun\ had lower than expected rotational velocities when compared to field stars in the same mass range and in similar evolutionary stages. This discrepancy suggested that they were in fact the evolved counterparts of less massive stars and, mass determinations for two of them were shown to be erroneous. We should note that, besides the studies quoted by \cite{lloyd11}, \cite{ghezzi10b} also showed that HD 154857 has \mstar\ = 1.21 \msun\ using a similar method as the one described in this work. HD 102956, on the other hand, was analyzed here and it is indeed massive (1.67 $\pm$ 0.10 \msun). It has \logg = 3.38 $\pm$ 0.16 and \vsini\ = 1.6 $\pm$ 0.6 km s$^{-1}$, thus being consistent with both planet hosting and field stars in the lower panel of Figure 1 from \cite{lloyd11}. Unfortunately, HD 190228 was not analyzed in our study, but recent determinations confirm its a lower mass star with 1.18 \msun\ \citep{jofre15}.}

\begin{figure}
\plotone{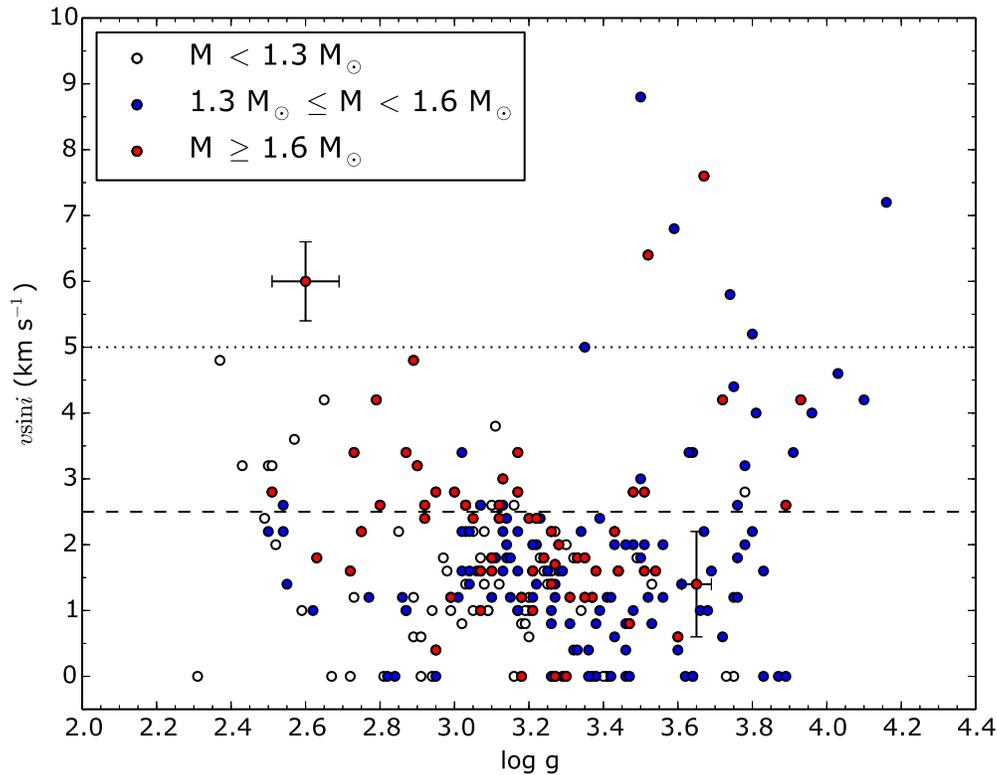}
\caption{\textbf{Rotational velocities \vsini\ for the subgiants as a function of \logg. Stars with masses \mstar $<$ 1.3 M$_{\odot}$, 1.3 M$_{\odot}$ $\leq$ \mstar $<$ 1.6 M$_{\odot}$ and \mstar $\geq$ 1.6 M$_{\odot}$ (i.e., Retired A Stars) are represented by open, blue and red circles, respectively. The black dashed and dotted lines show the limits \vsini\ = 3 and 5 km s$^{-1}$ discussed in the text. Stars with error bars are discussed in the text.}}
\label{plot_vsini_logg_mass}
\end{figure}

\textbf{In order to address this concern regarding the actual sample of Retired A Stars (note that \citealt{lloyd11} did not use the sample from \citealt{johnson10a}), we analyzed their rotational velocities (see Section \ref{vsini}) as a function of their masses (see Section \ref{evol_par}) and evolutionary stages (represented by \logg; see Section \ref{atm_par}) in Figure \ref{plot_vsini_logg_mass}. We can observe a very good agreement with the field stars shown in Figure 1 of \cite{lloyd11}. In the lower panel of this figure, no stars with \logg\ $\gtrsim$ 3.6 have \vsini\ $\lesssim$ 2.5 km s$^{-1}$. In our sample, only one star is found in this region: HD 171264. It has \logg\ = 3.65 $\pm$ 0.04, \mstar\ = 1.64 $\pm$ 0.08 \msun\ and \vsini\ = 1.4 $\pm$ 0.8 km s$^{-1}$. Thus, it is consistent with the field stars from \cite{lloyd11} within the uncertainties. We should also note that the value of \vsini\ = 2.4 km s$^{-1}$ from \cite{brewer16} is consistent with the lower limit determined by the field stars. This sample also seems to have an upper limit of $\approx$5 km s$^{-1}$ for stars with \logg\ $\lesssim$ 3.4. One star in our sample appears above this limit: HD 150331. It is a G0/2 bright giant (see Section \ref{atm_par}) with a mass of 2.05 \msun\ and an age of 1 Gyr (see Table \ref{evol_par}). In conclusion, the rotational velocities provide another confirmation that masses of the Retired A Stars were not overestimated.}

\subsection{Giant Planet-Metallicity-Mass Correlation}
\label{correlation}

In the previous sections, we showed that the atmospheric and evolutionary parameters derived for the subgiants are reliable. We now use these results to investigate how the formation of giant planets is influenced by stellar mass and metallicity.  We follow the method described in \cite{johnson10a} or \cite{montet14} and adopt a similar function to describe the fraction of stars with giant planets:

\begin{equation}
f(M_{\star},F) = CM_{\star}^{\alpha}10^{\beta F}
\label{planet_fraction}
\end{equation}

In the above equation, \mstar\ is the stellar mass in M$_{\odot}$ and $F \equiv$ [Fe/H] is the stellar metallicity. The free parameters $\alpha$, $\beta$ and $C$ are determined so the likelihood conditioned on the data is maximized. We decided to use uniform priors within the following intervals: $\alpha$ = (0.00, 3.00) with 0.05 steps, $\beta$ = (0.00, 3.00) with 0.05 steps and $C$ = (0.005, 0.150) with 0.005 steps. 

\textbf{The first test consists in our best effort to mimic as closely as possible the analysis done by \cite{johnson10a} and check if we are able to recover their results: $\alpha$ = 1.00, $\beta$ = 1.20 and $C$ = 0.07.} The respective confidence intervals are: (0.70, 1.30), (1.00, 1.40) and (0.06, 0.08). As we can see in the first line of Table \ref{results_cab}, our results are consistent with these values within the confidence intervals.

We then used the sample of FGKM dwarfs of \cite{johnson10a}, but replaced their subgiants with our own sample. It should be noted that this was not just an update of the parameters. The sample of subgiants in this work is not the same as the one from \cite{johnson10a}, there are 56 different objects. Some stars were included because they now fulfill the requirement of at least nine observations, while others were removed due to scientific reasons (e.g., confirmation that they are binaries or poor RV measurements due to enhanced activity or variability). 

Since we are only interested in the giant planets in this work, we considered stars that have detected planets with real or minimum masses larger than 0.50 M$_{Jup}$. This is almost twice Saturn's mass and close to the detection limit of 0.44 M$_{Jup}$ for a 0.4 M$_{\odot}$ star described in \cite{johnson10a}. If a star only has planets less massive than this limit, we treat it as if it does not have planets in the following analyses. We also used this lower limit to update the list of planet hosts in the FGKM sample. 

With these constraints, we obtain a new giant planet occurrence relation and see that our overall occurrence parameter, $C$, increases, which is expected since, as our mass threshold is lower and time baseline longer than \cite{johnson10a}, there are more detected planets in our revised sample. The values of $\alpha$ and $\beta$ change by the same amount, in opposite directions, but within the errors in both cases (see the second line of Table \ref{results_cab}). 

\begin{figure}
\plotone{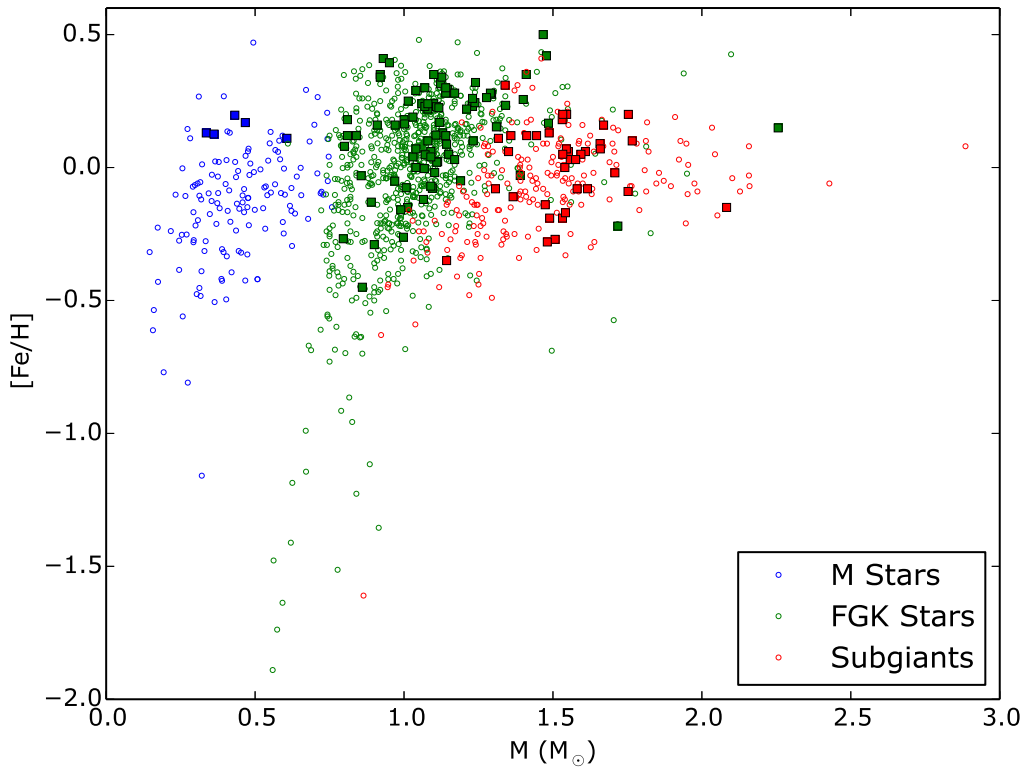}
\caption{Masses and metallicities for our complete sample of 1225 M (blue), FGK (green) and subgiants (red). Filled squares and open circles represent stars with and without planets, respectively.}
\label{plot_mass_met}
\end{figure}

Still using our results for the subgiants, we replaced the parameters and uncertainties of the M stars by those from \cite{montet14}. Whenever available, we also adopted parameters and errors from \cite{brewer16} for the FGK dwarfs, keeping the results from \cite{johnson10a} otherwise. This is our complete and updated sample of 1225 stars and it is depicted in Figure \ref{plot_mass_met}. Although there are some stars with very low metallicities, we highlight that there are no selection biases that cause a correlation between these quantity and the stellar masses. A linear fit to the entire set of stars returns a correlation coefficient $R^{2}$ = 0.08. When this complete sample is used, we notice slight changes in both $\alpha$ and $\beta$, but again the results are consistent with the previous values within the uncertainties (see the third line of Table \ref{results_cab} and Eq. \ref{planet_fraction_best}).

\cite{sw13} mention that a systematic offset as small as $\approx$+0.05 in the metallicities of the Retired A Stars could explain the larger occurrence rate of giant planets around them. In fact, our metallicities are, on average, 0.04 dex smaller than those of \cite{johnson10a} and \cite{brewer16} (see Section \ref{test_spocs}). Despite this, we can still see an increased occurrence rate of giant planets around more massive stars. 

We saw in Section \ref{test_spec_feh} that there is no reason to believe that our [Fe/H] values would be systematically higher for evolved stars relative to dwarfs. However, for completeness, we decided to test whether corrections to the metallicity scale (-0.05 and -0.10 dex) would change our results. We can see in Table \ref{results_cab} that the dependence of the occurrence rate of giant planets on metallicity decreases. As a consequence, the dependence on mass becomes increasingly stronger. Thus, if the Retired A Stars were more metal-poor, then planet formation would be a steeper function of stellar mass in order to explain the giant planets detected at the high-mass end of our sample.

We saw in Section \ref{mass_uvw} that a decrease of only 0.12 M$_{\odot}$ in the masses of the Retired A Stars is sufficient to reconcile their kinematics with the velocity dispersions of a sample of A5-F0 dwarfs. As this offset is almost equal to the typical uncertainties in our mass determinations, we tested how it affects the occurrence rate. Table \ref{results_cab} shows that the effect of this mass offest is negligible. Therefore, the correlation between giant planets, metallicity and mass is robust against the typical errors in the latter.

As discussed in Section \ref{offsets_atm_par}, a decrease in [Fe/H] will cause a decrease in mass, so these parameters are not independent. In order to account for this, we tested how giant planet occurrence is affected by simultaneous offsets in the metallicities (-0.05 and -0.10 dex) and masses (-0.12 M$_{\odot}$) of the subgiants. The results in Table \ref{results_cab} reveal a significant decrease in $\beta$ and an increase in $\alpha$, in a similar manner as observed when only [Fe/H] is altered.

\begin{figure}
\plotone{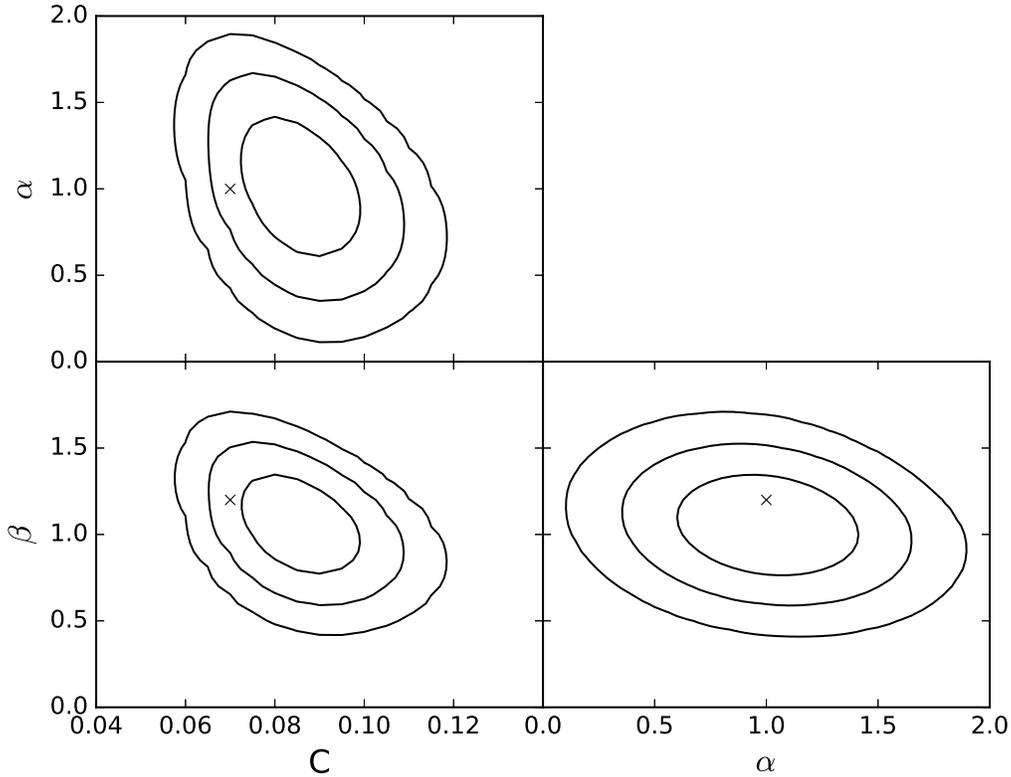}
\caption{Marginal posterior probability density functions for the parameters $\alpha$, $\beta$ and $C$ conditioned on the data. Lines represent the 68\%, 95\% and 99.7\% contours (from smaller to larger). The crosses represent the best-fitting values from \cite{johnson10a}: $\alpha$ = 1.00, $\beta$ = 1.20 and $C$ = 0.07.}
\label{plot_alpha_beta}
\end{figure}

These results confirm that giant planet occurrence is a clear function of both metallicity and mass and was not artificially produced by uncertainties or systematic offsets in any or both of these parameters. The best functional form to describe this dependence is (see also Figure \ref{plot_alpha_beta}):

\begin{equation}
f(M_{\star},F) = 0.085_{-0.010}^{+0.008}M_{\star}^{1.05_{-0.24}^{+0.28}}10^{1.05_{-0.17}^{+0.21}F}
\label{planet_fraction_best}
\end{equation}

It is interesting to note that, in the previous tests, both $\alpha$ and $\beta$ always change in opposite directions, meaning that lower metallicities can be compensated by larger masses and vice-versa (as already suggested in \citealt{ghezzi10b}). Therefore, it seems that the formation of giant planets depends on the quantity \mstar $\times 10^{[Fe/H]}$. In order to test this, we now describe planet occurrence in the following way:

\begin{equation}
f(M_{\star},F) = C(M_{\star} \times 10^{F})^{\gamma}
\label{planet_fraction_2}
\end{equation}

\begin{figure}
\plotone{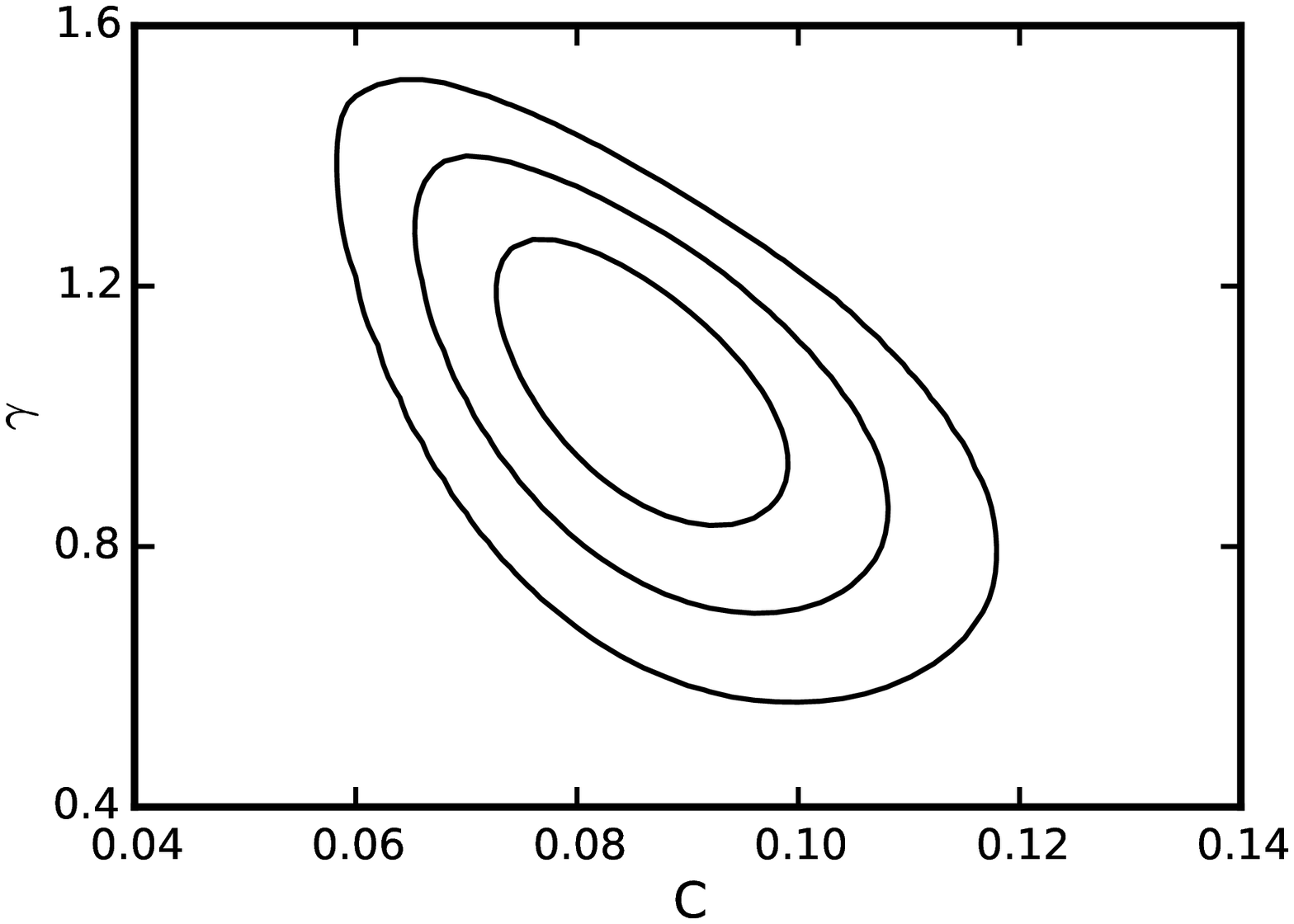}
\caption{Same as Figure \ref{plot_alpha_beta}, but now for parameters $\gamma$ and $C$.}
\label{plot_gamma}
\end{figure}

As before, the free parameters $\gamma$ and $C$ are determined so the likelihood conditioned on the data is maximized. We decided to use uniform priors within the following intervals: $\gamma$ = (0.00, 2.30) with 0.02 steps and $C$ = (0.002, 0.150) with 0.002 steps. Using the most updated results for all stars, we obtain the following best functional form to describe this dependence (see also Figure \ref{plot_gamma}):

\begin{equation}
f(M_{\star},F) = 0.086_{-0.010}^{+0.008}(M_{\star} \times 10^{F})^{1.04_{-0.16}^{+0.12}}
\label{planet_fraction_best_2}
\end{equation}

\begin{figure}
\plotone{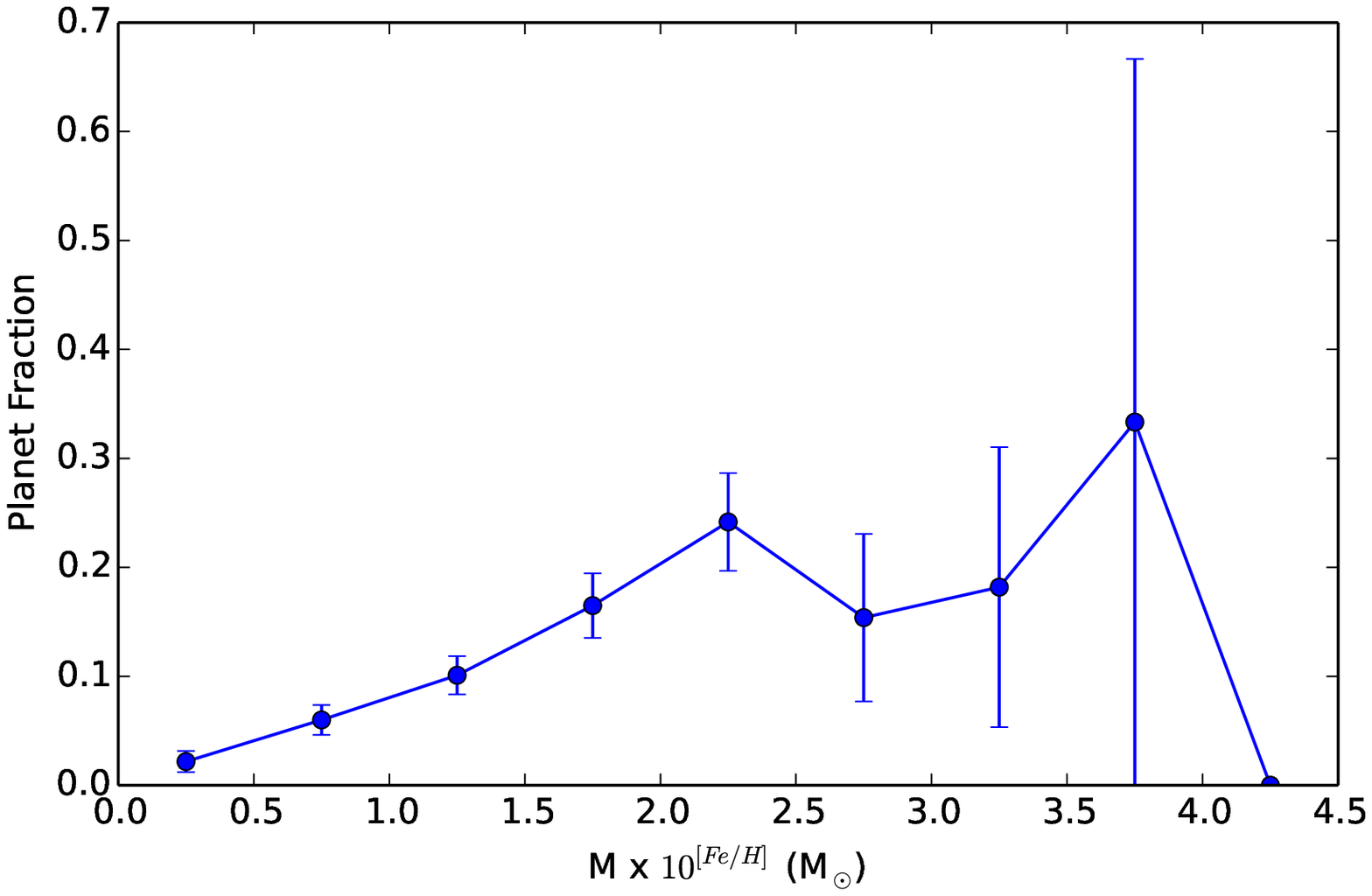}
\caption{Fraction of stars with planets as a function of the amount of metals in the protoplanetary disk.}
\label{plot_planet_fraction}
\end{figure}

We note that the best value for $\gamma$ is very close to 1.00. Offsets applied to the metallicities and/or masses of the subgiants did not change the results significantly (see Table \ref{results_cg}). If we remember that M$_{d} \propto$ \mstar, where M$_{d}$ and \mstar are the masses of the protoplanetary disk and stellar host, respectively \citep{andrews13}, the quantity \mstar $\times 10^{[Fe/H]}$ can be considered as proxy for the amount of metals in the disk from which planets formed. In this scenario, our results reveal that the formation of giant planets is an almost linear function of the amount of metals in the protoplanetary disk, going from $\approx$2\% in the interval 0.0 -- 0.5 M$_{\odot}$ to $\approx$24\% between 2.0 -- 2.5 M$_{\odot}$ (see Figure \ref{plot_planet_fraction}). The behavior beyond this interval is too uncertain due to low number statistics. Independently of the functional form chosen, our results provide a strong support for the core accretion hypothesis \citep{il04}. 

\section{Conclusions}
\label{conclusions}

We determined atmospheric (effective temperature, metallicity and surface gravity), rotational (\vsini), evolutionary (mass, radius and age) and kinematical ($UVW$ space velocities) parameters for a sample of 245 subgiants. Thorough tests and comparisons showed that our results agree well with those determined using other methods and exhibit acceptable offsets when systematic errors are applied to the input data. In particular, we estimate the typical uncertainties on the masses to be 0.11 M$_{\odot}$, while a maximum offset of 0.20 M$_{\odot}$ could be present if the input parameters were simultaneously affected by systematic errors. This larger value would correspond to an error of $\lesssim$ 13\%, much lower than the 50\% overestimate suggested by \cite{lloyd11,lloyd13}. 

We also presented a simple explanation to the apparent inconsistency between their masses and kinematics observed by \cite{sw13}. \textbf{By restricting our sample to Retired A Stars (i.e., stars with \mstar $\geq$ 1.6 M$_{\odot}$) and applying an offset of just -0.12 M$_{\odot}$, the velocity dispersions become consistent with the main-sequence A5-F0 stars from \cite{sw13}. If reddening is considered for the latter sample, this offset decreases to 0.04 \msun. The rotational velocities derived for our targets corroborated that the Retired A Stars are indeed massive, contrary to the concerns raised by \cite{lloyd11}. Therefore, a thorough and simultaneous analysis of kinematics, rotation and masses confirm that the latter were not significantly overestimated within the expected errors for the Retired A Stars.}    

Having confirmed the masses and metallicities derived in this study are reliable, we extended our sample with main-sequence FGKM stars from \cite{johnson10a}, but adopted updated parameters from \cite{montet14} and \cite{brewer16} (whenever available) and accounted for the most recent giant planet detections. This sample of 1225 stars was used to study the correlation between giant planet occurrence and both stellar mass and metallicity. We confirm recent results \citep[e.g.][]{johnson10a,reffert15,jones16} that show there is a higher probability of forming giant planets around more massive and metal-rich stars up to $\sim$2.0 -- 2.5 M$_{\odot}$. Tests with offsets in metallicity, mass or both confirm that this correlation is real, and not artificially produced by systematic uncertainties. 

Since mass and metallicity seem to be equally important for the formation of giant planets, we also studied planet occurrence as a function of $(M_* \times 10^{[Fe/H]} )^\gamma$. This quantity can be thought as the total amount of metals in the protoplanetary disk, assuming that M$_{d} \propto$ \mstar \citep{andrews13}. We observed that the formation of giant planets is an almost linear function of the total amount of metals in the protoplanetary disk ($\gamma \approx 1$). This result agrees with the suggestion given by \cite{ghezzi10b} and provides another strong evidence that most (if not all) giant planets form through the core accretion mechanism \citep[e.g.,][]{il04}. 

This study also highlights the crucial role detailed stellar characterization plays on a complete understanding of planet formation. In this sense, our team continues to follow-up and analyze samples of evolved stars with the goal of obtaining more accurate and precise models for the giant planet-mass-metallicity correlations. New planet detections and updated stellar parameters will be presented in future contributions. 

\clearpage

\acknowledgments

This work has made use of: SIMBAD database, operated at CDS, Strasbourg, France; VALD database, operated at Uppsala University, the Institute of Astronomy RAS in Moscow, and the University of Vienna; NASA's Astrophysics Data System (ADS) and Exoplanet Archive.

The authors would like to thank Leo Girardi for kindly sharing the PARAM code, Kevian Schlaufman for providing data for the discussion regarding the kinematics and Let\'icia Dutra-Ferreira for making the reduced spectra of the Hyades stars available. We also thank the referee for their insightful comments, which led to a much improved manuscript. LG would like to thank the financial support from Coordena\c c\~ao de Aperfei\c coamento de Pessoal de N\'ivel Superior (CAPES), Ci\^encia sem Fronteiras,
Harvard College Observatory, and Funda\c c\~ao Lemann. JAJ is grateful for the generous grant support provided by the Alfred P. Sloan and David \& Lucile Packard foundations.
Work by B.T.M. was performed in part under contract with the California Institute of Technology (Caltech)/Jet Propulsion Laboratory (JPL) funded by NASA through the Sagan Fellowship Program executed by the NASA Exoplanet Science Institute.



\facility{Keck (HIRES)}

\software{%
    numpy \citep{numpy},
    matplotlib \citep{matplotlib},
    MOOG \citep{sneden73}
    IRAF, PARAM, astrolibpy}

\clearpage

\startlongtable
\begin{deluxetable}{lrrrrrrrrr}
\tablecolumns{10}
\rotate
\tabletypesize{\scriptsize}
\tablewidth{0pt}
\tablecaption{{Sample of Subgiants}.
\label{table_sample}}
\tablehead{
\colhead{Star} & \colhead{$\alpha$} & \colhead{$\delta$} & \colhead{V} & \colhead{$A_{V}$} & \colhead{$\pi$} & 
\colhead{$\mu_{\alpha}$} & \colhead{$\mu_{\delta}$} & \colhead{RV} & \colhead{S/N} \\
\colhead{} & \colhead{(J2000.0)} & \colhead{(J2000.0)} & \colhead{} & \colhead{} & \colhead{(mas)} & 
\colhead{(mas yr$^{-1}$)} & \colhead{(mas yr$^{-1}$)} & \colhead{(km s$^{-1}$)} & \colhead{}
}
\startdata
HD 1100   & 003.81777959 & -06.13821196 &  7.762 $\pm$ 0.013 & 0.099 $\pm$ 0.035 &  7.12 $\pm$ 0.92 &  -36.51 &  -49.99 &   22.749 & 199 \\
HD 1293   & 004.27312298 & -01.65301289 &  8.376 $\pm$ 0.016 & 0.099 $\pm$ 0.035 &  6.31 $\pm$ 1.05 &   25.88 &   11.56 &   45.140 & 241 \\
HD 1384   & 004.55058081 & +52.41254980 &  8.097 $\pm$ 0.013 & 0.110 $\pm$ 0.015 &  9.34 $\pm$ 1.14 &  -14.90 &  -56.14 &  -35.185 & 144 \\
HD 1502   & 004.82110857 & +14.05475398 &  8.349 $\pm$ 0.015 & 0.108 $\pm$ 0.080 &  6.28 $\pm$ 0.75 &   74.64 &  -18.15 &  -10.024 & 170 \\
HD 2946   & 008.14629082 & -18.10912237 &  8.080 $\pm$ 0.013 & 0.099 $\pm$ 0.035 &  7.41 $\pm$ 0.94 &   34.13 &   15.59 &   10.460 & 254 \\
HD 3458   & 009.34561534 & -19.56502439 &  8.257 $\pm$ 0.013 & 0.099 $\pm$ 0.035 &  3.53 $\pm$ 0.80 &   -3.82 &   -3.96 &   -4.607 & 223 \\
HD 4313   & 011.41815996 & +07.84502103 &  7.829 $\pm$ 0.014 & 0.100 $\pm$ 0.074 &  7.30 $\pm$ 0.76 &   -5.14 &    6.69 &   14.482 & 181 \\
HD 4395   & 011.55728796 & -11.45237787 &  7.689 $\pm$ 0.013 & 0.099 $\pm$ 0.028 & 10.56 $\pm$ 0.68 & -106.18 &  -38.11 &   -0.439 & 249 \\
\enddata
\end{deluxetable}

\clearpage

\startlongtable
\begin{deluxetable}{cccccc}
\tablecolumns{6}
\tabletypesize{\normalsize}
\tablewidth{0pt}
\tablecaption{List of Fe I and Fe II lines.
\label{table_line_list}}
\tablehead{
\colhead{$\lambda$} & \colhead{Ion\tablenotemark{a}} & \colhead{$\chi$} &
\colhead{$\log\ gf$} & \colhead{$log\ \Gamma_{w}$\tablenotemark{b}} & \colhead{EW$_{\odot}$} \\
\colhead{(\AA)} & \colhead{} & \colhead{(eV)} & 
\colhead{} & \colhead{(4$\pi s N_{H})^{-1}$} & \colhead{(m\AA)}
}
\startdata
5023.185 & 26.0 & 4.283 & -1.524 & -7.135 & 37.9 \\
5025.303 & 26.0 & 4.284 & -1.919 & -7.140 & 20.6 \\
5044.211 & 26.0 & 2.851 & -2.206 & -7.280 & 72.3 \\
5054.642 & 26.0 & 3.640 & -2.087 & -7.599 & 38.3 \\
5058.496 & 26.0 & 3.642 & -2.809 & -7.599 & 12.1 \\
5067.149 & 26.0 & 4.220 & -1.068 & -7.187 & 65.3 \\
5109.651 & 26.0 & 4.301 & -0.853 & -7.150 & 74.2 \\
5159.057 & 26.0 & 4.283 & -0.932 & -7.175 & 70.1 \\
\enddata
\tablenotetext{a}{Fe I and Fe II ions are represented by the notations 26.0 and 26.1, following the format used by MOOG.}
\tablenotetext{b}{$log\ \Gamma_{w}$ is the logarithm of the van der Waals damping constant.}
\end{deluxetable}

\clearpage

\startlongtable
\begin{deluxetable}{lccrccccc}
\tablecolumns{9}
\tabletypesize{\scriptsize}
\tablewidth{0pt}
\tablecaption{Atmospheric Parameters.
\label{table_atm_par}}
\tablehead{
\colhead{Star} & \colhead{\teff} & \colhead{\logg} & \colhead{\feh} & \colhead{\vmic} & 
\colhead{$\sigma$} & \colhead{N} & \colhead{$\sigma$} & \colhead{N} \\
\colhead{} & \colhead{(K)} & \colhead{} & \colhead{} & \colhead{(km\ s$^{-1}$)} &
\colhead{(Fe I)} & \colhead{(Fe II)} & \colhead{(Fe I)} & \colhead{(Fe II)}
}
\startdata
HD 1100 	&	4808	$\pm$	29	&	2.89	$\pm$	0.17	&	-0.06	$\pm$	0.02	&	1.165	$\pm$	0.030	&	0.05	&	133	&	0.11	&	17	\\
HD 1293 	&	5160	$\pm$	21	&	3.27	$\pm$	0.06	&	-0.44	$\pm$	0.01	&	0.990	$\pm$	0.030	&	0.04	&	133	&	0.06	&	14	\\
HD 1384 	&	4875	$\pm$	50	&	3.39	$\pm$	0.18	&	0.31	$\pm$	0.02	&	1.093	$\pm$	0.060	&	0.10	&	136	&	0.10	&	11	\\
HD 1502 	&	4988	$\pm$	22	&	3.21	$\pm$	0.07	&	-0.00	$\pm$	0.01	&	1.123	$\pm$	0.030	&	0.05	&	137	&	0.05	&	15	\\
HD 2946 	&	5488	$\pm$	15	&	3.64	$\pm$	0.06	&	-0.23	$\pm$	0.01	&	1.171	$\pm$	0.020	&	0.03	&	138	&	0.06	&	15	\\
HD 3458 	&	5168	$\pm$	25	&	2.90	$\pm$	0.09	&	-0.04	$\pm$	0.02	&	1.420	$\pm$	0.030	&	0.05	&	140	&	0.06	&	14	\\
HD 4313 	&	4943	$\pm$	29	&	3.26	$\pm$	0.06	&	0.09	$\pm$	0.02	&	1.104	$\pm$	0.030	&	0.06	&	139	&	0.06	&	14	\\
HD 4395 	&	5610	$\pm$	18	&	3.75	$\pm$	0.07	&	-0.15	$\pm$	0.01	&	1.218	$\pm$	0.030	&	0.04	&	145	&	0.06	&	16	\\
\enddata
\end{deluxetable}

\clearpage

\begin{deluxetable}{ccccc}
\tablecolumns{5}
\tabletypesize{\normalsize}
\tablewidth{0pt}
\tablecaption{Atmospheric Parameters Sensitivities.
\label{atm_par_sensitivities}}
\tablehead{
\colhead{Parameter} & \colhead{$\Delta$\teff} & \colhead{$\Delta$\logg} & \colhead{$\Delta$\feh} & \colhead{$\Delta$\vmic} \\
\colhead{} & \colhead{($\pm$100 K)} & \colhead{($\pm$0.20 dex)} & \colhead{($\pm$0.20 dex)} & \colhead{($\pm$0.200 km s$^{-1}$)}
}
\startdata
\teff\ (K)            &              \nodata &       $^{-79}_{+29}$ &       $^{+29}_{-30}$ &     $^{+72}_{-37}$ \\
\logg\ (dex)          &   $^{+0.27}_{-0.22}$ &              \nodata &   $^{-0.05}_{+0.02}$ & $^{+0.20}_{-0.09}$ \\
\feh\  (dex)          &   $^{+0.08}_{-0.07}$ &   $^{+0.01}_{-0.01}$ &              \nodata & $^{-0.02}_{+0.05}$ \\
\vmic\ (km s$^{-1}$)  & $^{-0.044}_{-0.009}$ & $^{-0.098}_{+0.043}$ & $^{+0.020}_{-0.024}$ &            \nodata \\
\enddata
\end{deluxetable} 

\clearpage

\startlongtable
\begin{deluxetable}{lccc}
\tablecolumns{4}
\tabletypesize{\normalsize}
\tablewidth{0pt}
\tablecaption{Broadening Parameters.
\label{table_vsini}}
\tablehead{
\colhead{Star} & \colhead{FWHM$_{inst}$} & \colhead{$V_{macro}$} & \colhead{\vsini} \\
\colhead{} & \colhead{(\AA)} & \colhead{(km s$^{-1}$)} & \colhead{(km s$^{-1}$)}
}
\startdata
HD 1100    & 0.112 & 4.00 &  0.6 $\pm$ 0.8 \\
HD 1293    & 0.112 & 3.51 &  2.2 $\pm$ 0.6 \\
HD 1384    & 0.112 & 3.39 &  2.4 $\pm$ 0.4 \\
HD 1502    & 0.112 & 3.43 &  1.2 $\pm$ 1.0 \\
HD 2946    & 0.112 & 3.82 &  3.4 $\pm$ 0.8 \\
HD 3458    & 0.112 & 4.00 &  3.2 $\pm$ 0.6 \\
HD 4313    & 0.112 & 3.41 &  2.2 $\pm$ 0.4 \\
HD 4395    & 0.112 & 4.03 &  1.2 $\pm$ 1.0 \\
\enddata
\end{deluxetable}

\clearpage

\startlongtable
\begin{deluxetable}{lrrrr}
\tablecolumns{5}
\tabletypesize{\scriptsize}
\tablewidth{0pt}
\tablecaption{Evolutionary Parameters.
\label{table_evol_par}}
\tablehead{
\colhead{Star} & \colhead{M} & \colhead{R} & \colhead{\logg} & \colhead{Age} \\ 
\colhead{} & \colhead{(\msun)} & \colhead{(\rsun)} & \colhead{} & \colhead{(Gyr)}
}
\startdata
HD 1100   & 1.284 $\pm$ 0.112 &  5.627 $\pm$ 0.648 & 3.018 $\pm$ 0.075 &  4.449 $\pm$ 1.303 \\
HD 1293   & 1.250 $\pm$ 0.099 &  3.674 $\pm$ 0.502 & 3.377 $\pm$ 0.088 &  3.672 $\pm$ 0.929 \\
HD 1384   & 1.381 $\pm$ 0.105 &  3.581 $\pm$ 0.449 & 3.443 $\pm$ 0.092 &  4.067 $\pm$ 0.935 \\
HD 1502   & 1.540 $\pm$ 0.097 &  4.355 $\pm$ 0.532 & 3.319 $\pm$ 0.081 &  2.617 $\pm$ 0.479 \\
HD 2946   & 1.452 $\pm$ 0.111 &  3.221 $\pm$ 0.471 & 3.556 $\pm$ 0.092 &  2.638 $\pm$ 0.566 \\
HD 3458   & 2.102 $\pm$ 0.257 &  6.393 $\pm$ 1.469 & 3.121 $\pm$ 0.147 &  1.019 $\pm$ 0.356 \\
HD 4313   & 1.658 $\pm$ 0.109 &  5.068 $\pm$ 0.533 & 3.220 $\pm$ 0.069 &  2.188 $\pm$ 0.430 \\
HD 4395   & 1.358 $\pm$ 0.047 &  2.681 $\pm$ 0.191 & 3.681 $\pm$ 0.050 &  3.298 $\pm$ 0.408 \\
\enddata
\end{deluxetable}

\clearpage

\startlongtable
\begin{deluxetable}{lrrr}
\tablecolumns{4}
\tabletypesize{\normalsize}
\tablewidth{0pt}
\tablecaption{Galactic \textit{UVW} space velocities.
\label{table_kinematics}}
\tablehead{
\colhead{Star} & \colhead{$U$} & \colhead{$V$} & \colhead{$W$} \\ 
\colhead{} & \colhead{(km s$^{-1}$)} & \colhead{(km s$^{-1}$)} & \colhead{(km s$^{-1}$)}
}
\startdata
HD 1100   &   36.132 &   -6.115 &  -29.551 \\
HD 1293   &  -25.861 &   17.024 &  -39.147 \\
HD 1384   &   28.312 &  -29.649 &  -20.650 \\
HD 1502   &  -40.350 &  -41.970 &   -8.545 \\
HD 2946   &  -24.045 &   -0.934 &  -10.325 \\
HD 3458   &    7.253 &   -2.470 &    4.137 \\
HD 4313   &   -3.335 &   11.923 &   -9.298 \\
HD 4395   &   49.210 &   11.520 &   -3.237 \\
\enddata
\end{deluxetable}

\clearpage

\startlongtable
\begin{deluxetable}{lcr}
\tablecolumns{3}
\tabletypesize{\normalsize}
\tablewidth{0pt}
\tablecaption{K magnitudes and Photometric Effective Temperatures.
\label{table_phot_teff}}
\tablehead{
\colhead{Star} & \colhead{$K_{S}$} & \colhead{$T_{eff}^{Phot}$} \\ 
\colhead{} & \colhead{} & \colhead{(K)}
}
\startdata
HD 1100   &  5.319 & 4801 $\pm$  32 \\
HD 1293   &  6.354 & 5247 $\pm$  46 \\
HD 1384   &  5.785 & 4947 $\pm$  37 \\
HD 1502   &  6.120 & 5027 $\pm$  44 \\
HD 2946   &  6.275 & 5508 $\pm$  45 \\
HD 3458   &  6.180 & 5156 $\pm$  31 \\
HD 4313   &  5.595 & 5015 $\pm$  40 \\
HD 4395   &  5.972 & 5620 $\pm$  43 \\
\enddata
\end{deluxetable}

\clearpage

\begin{deluxetable}{crrcccccccc}
\tablecolumns{11}
\tabletypesize{\scriptsize}
\tablewidth{0pt}
\tablecaption{Results for the Giant Plant Occurrence given by Equation \ref{planet_fraction}.
\label{results_cab}}
\tablehead{
\colhead{Retired A Stars} & \colhead{$\Delta [Fe/H]$} & \colhead{$\Delta M$} & \colhead{FGK Dwarfs} & \colhead{M Dwarfs} & \colhead{$\alpha$} & \colhead{Confidence} & \colhead{$\beta$} & \colhead{Confidence} & \colhead{C} & \colhead{Confidence} \\
\colhead{} & \colhead{(dex)} & \colhead{(M$_{\odot}$)} & \colhead{} & \colhead{} & \colhead{} & \colhead{Interval} & \colhead{} & \colhead{Interval} & \colhead{} & \colhead{Interval}
}
\startdata
J10        &    0.00 &    0.00 & J10     & J10 & 0.99 & (0.79,1.29) & 1.19 & (0.95,1.25) & 0.072 & (0.064,0.081) \\
This work  &    0.00 &    0.00 & J10     & J10 & 1.10 & (0.92,1.40) & 1.00 & (0.81,1.12) & 0.085 & (0.073,0.091) \\
This work  &    0.00 &    0.00 & J10,B16 & M14 & 1.05 & (0.81,1.33) & 1.05 & (0.88,1.26) & 0.085 & (0.075,0.093) \\
This work  & $-$0.05 &    0.00 & J10,B16 & M14 & 1.10 & (0.87,1.39) & 0.95 & (0.81,1.17) & 0.090 & (0.078,0.095) \\
This work  & $-$0.10 &    0.00 & J10,B16 & M14 & 1.25 & (0.97,1.49) & 0.90 & (0.74,1.08) & 0.090 & (0.079,0.098) \\
This work  &    0.00 & $-$0.12 & J10,B16 & M14 & 1.05 & (0.79,1.33) & 1.05 & (0.86,1.24) & 0.085 & (0.077,0.095) \\
This work  & $-$0.05 & $-$0.12 & J10,B16 & M14 & 1.15 & (0.88,1.42) & 0.95 & (0.78,1.14) & 0.090 & (0.080,0.098) \\
This work  & $-$0.10 & $-$0.12 & J10,B16 & M14 & 1.15 & (0.92,1.46) & 0.85 & (0.71,1.06) & 0.095 & (0.082,0.100) \\
\enddata
\tablerefs{J10: \cite{johnson10a}; M14: \cite{montet14}; B16: \cite{brewer16}.}
\end{deluxetable} 

\clearpage

\begin{deluxetable}{crrcccccc}
\tablecolumns{9}
\tabletypesize{\scriptsize}
\tablewidth{0pt}
\tablecaption{Results for the Giant Plant Occurrence given by Equation \ref{planet_fraction_2}.
\label{results_cg}}
\tablehead{
\colhead{Retired A Stars} & \colhead{$\Delta [Fe/H]$} & \colhead{$\Delta M$} & \colhead{FGK Dwarfs} & \colhead{M Dwarfs} & \colhead{$\gamma$} & \colhead{Confidence} & \colhead{C} & \colhead{Confidence} \\
\colhead{} & \colhead{(dex)} & \colhead{(M$_{\odot}$)} & \colhead{} & \colhead{} & \colhead{} & \colhead{Interval} & \colhead{} & \colhead{Interval}
}
\startdata
This work  &    0.00 &    0.00 & J10,B16 & M14 & 1.04 & (0.88,1.16) & 0.086 & (0.076,0.094) \\
This work  & $-$0.05 &    0.00 & J10,B16 & M14 & 1.04 & (0.87,1.16) & 0.088 & (0.079,0.097) \\
This work  & $-$0.10 &    0.00 & J10,B16 & M14 & 1.00 & (0.84,1.12) & 0.092 & (0.082,0.100) \\
This work  &    0.00 & $-$0.12 & J10,B16 & M14 & 1.04 & (0.87,1.15) & 0.088 & (0.078,0.096) \\
This work  & $-$0.05 & $-$0.12 & J10,B16 & M14 & 1.02 & (0.84,1.13) & 0.090 & (0.081,0.099) \\
This work  & $-$0.10 & $-$0.12 & J10,B16 & M14 & 0.96 & (0.81,1.09) & 0.094 & (0.084,0.102) \\
\enddata
\tablerefs{J10: \cite{johnson10a}; M14: \cite{montet14}; B16: \cite{brewer16}.}
\end{deluxetable} 


\begin{thebibliography}{}

\bibitem[Anderson \& Francis(2012)]{af12} Anderson, E., \& Francis, C.\ 2012, Astronomy Letters, 38, 331

\bibitem[Andrews et al.(2013)]{andrews13} Andrews, S.~M., Rosenfeld, K.~A., Kraus, A.~L., \& Wilner, D.~J.\ 2013, \apj, 771, 129

\bibitem[Arenou et al.(1992)]{arenou92} Arenou, F., Grenon, M., \& Gomez, A.\ 1992, \aap, 258, 104 

\bibitem[Asplund et al.(2009)]{asplund09} Asplund, M., Grevesse, N., Sauval, A.~J., \& Scott, P.\ 2009, \araa, 47, 481

\bibitem[Barklem \& Aspelund-Johansson(2005)]{baj05} Barklem, P.~S., \& Aspelund-Johansson, J.\ 2005, \aap, 435, 373 


\bibitem[Barklem et al.(2000)]{barklem00} Barklem, P.~S., Piskunov, N., \& O'Mara, B.~J.\ 2000, \aaps, 142, 467

\bibitem[Bedell et al.(2014)]{bedell14} Bedell, M., Mel{\'e}ndez, J., Bean, J.~L., et al.\ 2014, \apj, 795, 23

\bibitem[Bressan et al.(2012)]{bressan12} Bressan, A., Marigo, P., Girardi, L., et al.\ 2012, \mnras, 427, 127

\bibitem[Brewer et al.(2016)]{brewer16} Brewer, J.~M., Fischer, D.~A., Valenti, J.~A., \& Piskunov, N.\ 2016, \apjs, 225, 32

\bibitem[Butler et al.(2006)]{butler06} Butler, R.~P., Johnson, J.~A., Marcy, G.~W., et al.\ 2006, \pasp, 118, 1685

\bibitem[Butler et al.(1996)]{butler96} Butler, R.~P., Marcy, G.~W., Williams, E., et al.\ 1996, \pasp, 108, 500

\bibitem[Campante et al.(2017)]{campante17} Campante, T.~L., Veras, D., North, T.~S.~H., et al.\ 2017, \mnras, 469, 1360 

\bibitem[Carroll \& Ostlie(2006)]{co06} Carroll, B.~W., \& Ostlie, D.~A.\ 2006, An Introduction to Modern Astrophysics, 2nd edition. (San Francisco, CA: Pearson, Addison-Wesley)

\bibitem[Casagrande et al.(2010)]{casagrande10} Casagrande, L., Ram{\'{\i}}rez, I., Mel{\'e}ndez, J., Bessell, M., \& Asplund, M.\ 2010, \aap, 512, A54

\bibitem[Casagrande et al.(2011)]{casagrande11} Casagrande, L., Sch{\"o}nrich, R., Asplund, M., et al.\ 2011, \aap, 530, A138

\bibitem[Castelli \& Kurucz(2004)]{ck04} Castelli, F., \& Kurucz, R. L. 2004, in IAU Symp. 210, Modelling of Stellar Atmospheres, ed. N. Piskunov, et al. (Dordrecht: Kluwer), poster A20 (arXiv:astro-ph/0405087)

\bibitem[Chabrier(2001)]{chabrier01} Chabrier, G.\ 2001, \apj, 554, 1274 

\bibitem[Chubak et al.(2012)]{chubak12} Chubak, C., Marcy, G., Fischer, D.~A., et al.\ 2012, arXiv:1207.6212

\bibitem[Cutri et al.(2003)]{cutri2003} Cutri, R. M., et al. 2003, The IRSA 2MASS All-Sky Point Source Catalog, NASA/IPAC Infrared Science Archive

\bibitem[da Silva et al.(2006)]{dasilva06} da Silva, L., Girardi, L., Pasquini, L., et al.\ 2006, \aap, 458, 609

\bibitem[Demarque et al.(2004)]{demarque04} Demarque, P., Woo, J.-H., Kim, Y.-C., \& Yi, S.~K.\ 2004, \apjs, 155, 667

\bibitem[Dutra-Ferreira et al.(2016)]{df16} Dutra-Ferreira, L., Pasquini, L., Smiljanic, R., Porto de Mello, G.~F., \& Steffen, M.\ 2016, \aap, 585, A75

\bibitem[ESA(1997)]{esa97} ESA 1997, The Hipparcos and Tycho Catalogues, ESA Special Publication, 1200 (Noordwijk: ESA)

\bibitem[Fischer \& Valenti(2005)]{fv05} Fischer, D.~A., \& Valenti, J.~A.\ 2005, \apj, 622, 1102

\bibitem[Fressin et al.(2013)]{fressin13} Fressin, F., Torres, G., Charbonneau, D., et al.\ 2013, \apj, 766, 81

\bibitem[Gaidos et al.(2013)]{gaidos13} Gaidos, E., Fischer, D.~A., Mann, A.~W., \& Howard, A.~W.\ 2013, \apj, 771, 18 

\bibitem[Ghezzi et al.(2010a)]{ghezzi10a} Ghezzi, L., Cunha, K., Smith, V.~V., de Ara\'ujo, F.~X., Schuler, S., \& de la Reza, R.\ 2010, \apj, 720, 1290

\bibitem[Ghezzi et al.(2010b)]{ghezzi10b} Ghezzi, L., Cunha, K., Schuler, S.~C., \& Smith, V.~V.\ 2010, \apj, 725, 721

\bibitem[Ghezzi \& Johnson(2015)]{gj15} Ghezzi, L., \& Johnson, J.~A.\ 2015, \apj, 812, 96

\bibitem[Girardi et al.(2005)]{girardi05} Girardi, L., Groenewegen, M.~A.~T., Hatziminaoglou, E., \& da Costa, L.\ 2005, \aap, 436, 895

\bibitem[Gontcharov(2006)]{gontcharov06} Gontcharov, G.~A.\ 2006, Astronomy Letters, 32, 759

\bibitem[Gontcharov \& Mosenkov(2017)]{gontcharov17} Gontcharov, G.~A., \& Mosenkov, A.~V.\ 2017, \mnras, 472, 3805

\bibitem[Gonzalez(1997)]{gonzalez97} Gonzalez, G.\ 1997, \mnras, 285, 403

\bibitem[Gonz{\'a}lez Hern{\'a}ndez \& Bonifacio(2009)]{gh09} Gonz{\'a}lez Hern{\'a}ndez, J.~I., \& Bonifacio, P.\ 2009, \aap, 497, 497

\bibitem[Hakkila et al.(1997)]{hakkila97} Hakkila, J., Myers, J.~M., Stidham, B.~J., \& Hartmann, D.~H.\ 1997, \aj, 114, 2043

\bibitem[Hinkel et al.(2016)]{hinkel16} Hinkel, N.~R., Young, P.~A., Pagano, M.~D., et al.\ 2016, \apjs, 226, 4

\bibitem[H{\o}g et al.(2000)]{hog00} H{\o}g, E., Fabricius, C., Makarov, V.~V., et al.\ 2000, \aap, 355, L27

\bibitem[{Hunter {et~al.}(2007)}]{matplotlib}
Hunter, J.~D., {et~al.} 2007, Computing in science and engineering, 9, 90

\bibitem[Ida \& Lin(2004)]{il04} Ida, S., \& Lin, D.~N.~C.\ 2004, \apj, 616, 567

\bibitem[Jofr{\'e} et al.(2015)]{jofre15} Jofr{\'e}, E., Petrucci, R., Saffe, C., et al.\ 2015, \aap, 574, A50

\bibitem[Johnson et al.(2006)]{johnson06} Johnson, J.~A., Marcy, G.~W., Fischer, D.~A., et al.\ 2006, \apj, 652, 1724 

\bibitem[Johnson et al.(2010a)]{johnson10a} Johnson, J.~A., Aller, K.~M., Howard, A.~W., \& Crepp, J.~R.\ 2010, \pasp, 122, 905

\bibitem[Johnson et al.(2010b)]{johnson10b} Johnson, J.~A., Howard, A.~W., Bowler, B.~P., et al.\ 2010, \pasp, 122, 701

\bibitem[Johnson et al.(2014)]{johnson14} Johnson, J.~A., Huber, D., Boyajian, T., et al.\ 2014, \apj, 794, 15  

\bibitem[Johnson & Wright (2013)]{johnsonwright13} Johnson
\& Wright, J.~T.\ 2013, arXiv:1306.6627v1

\bibitem[Johnson et al.(2013)]{johnson13} Johnson, J.~A., Morton, T.~D., \& Wright, J.~T.\ 2013, \apj, 763, 53

\bibitem[Jones et al.(2016)]{jones16} Jones, M.~I., Jenkins, J.~S., Brahm, R., et al.\ 2016, \aap, 590, A38

\bibitem[Kraft(1967)]{kraft67} Kraft, R.~P.\ 1967, \apj, 150, 551 

\bibitem[Kurucz et al.(1984)]{kurucz84} Kurucz, R.~L., Furenlid, I., Brault, J., \& Testerman, L.\ 1984, Solar Flux Atlas from 296 to 1300 nm (Cambridge: Harvard Univ. Press)

\bibitem[Liu et al.(2014)]{liu14} Liu, F., Asplund, M., Ram{\'{\i}}rez, I., Yong, D., \& Mel{\'e}ndez, J.\ 2014, \mnras, 442, L51 

\bibitem[Lloyd(2011)]{lloyd11} Lloyd, J.~P.\ 2011, \apjl, 739, L49 

\bibitem[Lloyd(2013)]{lloyd13} Lloyd, J.~P.\ 2013, \apjl, 774, L2

\bibitem[Mashonkina et al.(2011)]{mashonkina11} Mashonkina, L., Gehren, T., Shi, J.-R., Korn, A.~J., \& Grupp, F.\ 2011, \aap, 528, A87

\bibitem[Massarotti et al.(2008)]{massarotti08} Massarotti, A., Latham, D.~W., Stefanik, R.~P., \& Fogel, J.\ 2008, \aj, 135, 209

\bibitem[Montet et al.(2014)]{montet14} Montet, B.~T., Crepp, J.~R., Johnson, J.~A., Howard, A.~W., \& Marcy, G.~W.\ 2014, \apj, 781, 28 

\bibitem[Niedzielski et al.(2015)]{niedzielski15} Niedzielski, A., Villaver, E., Wolszczan, A., et al.\ 2015, \aap, 573, A36

\bibitem[Nordstr{\"o}m et al.(2004)]{nordstrom04} Nordstr{\"o}m, B., Mayor, M., Andersen, J., et al.\ 2004, \aap, 418, 989

\bibitem[North et al.(2017)]{north17} North, T.~S.~H., Campante, T.~L., Miglio, A., et al.\ 2017, \mnras, 472, 1866

\bibitem[Pavlenko et al.(2012)]{pavlenko12} Pavlenko, Y.~V., Jenkins, J.~S., Jones, H.~R.~A., Ivanyuk, O., \& Pinfield, D.~J.\ 2012, \mnras, 422, 542

\bibitem[Ram{\'{\i}}rez \& Mel{\'e}ndez(2005)]{rm05} Ram{\'{\i}}rez, I., \& Mel{\'e}ndez, J.\ 2005, \apj, 626, 465

\bibitem[Reffert et al.(2015)]{reffert15} Reffert, S., Bergmann, C., Quirrenbach, A., Trifonov, T., \& K{\"u}nstler, A.\ 2015, \aap, 574, A116

\bibitem[Robin et al.(2003)]{robin03} Robin, A.~C., Reyl{\'e}, C., Derri{\`e}re, S., \& Picaud, S.\ 2003, \aap, 409, 523

\bibitem[Ryabchikova et al.(2015)]{ryabchikova15} Ryabchikova, T., Piskunov, N., Kurucz, R.~L., et al.\ 2015, \physscr, 90, 054005

\bibitem[Schlaufman \& Winn(2013)]{sw13} Schlaufman, K.~C., \& Winn, J.~N.\ 2013, \apj, 772, 143

\bibitem[Schuler et al.(2011)]{schuler11} Schuler, S.~C., Flateau, D., Cunha, K., et al.\ 2011, \apj, 732, 55

\bibitem[Sneden(1973)]{sneden73} Sneden, C. 1973, PhD thesis, Univ. Texas, Austin

\bibitem[Soubiran et al.(2016)]{soubiran16} Soubiran, C., Le Campion, J.-F., Brouillet, N., \& Chemin, L.\ 2016, \aap, 591, A118 

\bibitem[Sousa et al.(2015a)]{sousa15a} Sousa, S.~G., Santos, N.~C., Adibekyan, V., Delgado-Mena, E., \& Israelian, G.\ 2015, \aap, 577, A67

\bibitem[Sousa et al.(2014)]{sousa14} Sousa, S.~G., Santos, N.~C., Adibekyan, V., et al.\ 2014, \aap, 561, A21

\bibitem[Sousa et al.(2008)]{sousa08} Sousa, S.~G., Santos, N.~C., Mayor, M., et al.\ 2008, \aap, 487, 373

\bibitem[Sousa et al.(2015b)]{sousa15b} Sousa, S.~G., Santos, N.~C., Mortier, A., et al.\ 2015, \aap, 576, A94

\bibitem[Stassun \& Torres(2016)]{st16} Stassun, K.~G., \& Torres, G.\ 2016, \apjl, 831, L6

\bibitem[Stassun et al.(2017)]{stassun17} Stassun, K.~G., Collins, K.~A., \& Gaudi, B.~S.\ 2017, \aj, 153, 136

\bibitem[Stello et al.(2017)]{stello17} Stello, D., Huber, D., Grundahl, F., et al.\ 2017, \mnras, 472, 4110 

\bibitem[Takeda \& Tajitsu(2015)]{tt15} Takeda, Y., \& Tajitsu, A.\ 2015, \mnras, 450, 397 

\bibitem[Torres et al.(2010)]{torres10} Torres, G., Andersen, J., \& Gim{\'e}nez, A.\ 2010, \aapr, 18, 67

\bibitem[Torres et al.(2012)]{torres12} Torres, G., Fischer, D.~A., Sozzetti, A., et al.\ 2012, \apj, 757, 161 

\bibitem[Tsantaki et al.(2013)]{tsantaki13} Tsantaki, M., Sousa, S.~G., Adibekyan, V.~Z., et al.\ 2013, \aap, 555, A150

\bibitem[Valenti \& Fischer(2005)]{vf05} Valenti, J.~A., \& Fischer, D.~A.\ 2005, \apjs, 159, 141

\bibitem[{Van Der~Walt {et~al.}(2011)Van Der~Walt, Colbert, \& Varoquaux}]{numpy} Van Der~Walt, S., Colbert, S.~C., \& Varoquaux, G. 2011, Computing in Science \& Engineering, 13, 22

\bibitem[van Leeuwen(2007)]{vanleeuwen07} van Leeuwen, F.\ 2007, \aap, 474, 653

\bibitem[van Saders \& Pinsonneault(2013)]{vsp13} van Saders, J.~L., \& Pinsonneault, M.~H.\ 2013, \apj, 776, 67 

\bibitem[Vogt et al.(1994)]{vogt94} Vogt, S.~S., Allen, S.~L., Bigelow, B.~C., et al.\ 1994, \procspie, 2198, 362

\bibitem[Wang \& Fischer(2015)]{wang15} Wang, J., \& Fischer, D.~A.\ 2015, \aj, 149, 14

\bibitem[Wright(2005)]{wright05} Wright, J.~T.\ 2005, \pasp, 117, 657 

\end{thebibliography}
\end{document}